# Geo-mechanical aspects for breakage detachment of rock fines by Darcy's flow

Abolfazl Hashemi[1], Sara Borazjani[1], Cuong Nguyen[1], Grace Loi[1], Nastaran Khazali[1], Alex Badalyan[1], Yutong Yang[1], Zhao Feng Tian[2], Heng Zheng Ting[2], Bryant Dang-Le[1], Thomas Russell[1], Pavel Bedrikovetsky[1]

[1]*School of Chemical Engineering, The University of Adelaide 5000, SA, Australia*

[2]*School of Mechanical Engineering, The University of Adelaide 5000, SA, Australia*

ABSTRACT    Suspension-colloidal-nano transport in porous media encompasses the detachment of detrital fines against electrostatic attraction and authigenic fines by breakage, from the rock surface. While much is currently known about the underlying mechanisms governing detachment of detrital particles, including detachment criteria at the pore scale and its upscaling for the core scale, a critical gap exists due to absence of this knowledge for authigenic fines. Integrating 3D Timoshenko's beam theory of elastic cylinder deformation with CFD-based model for viscous flow around the attached particle and with strength failure criteria for particle-rock bond, we developed a novel theory for fines detachment by breakage at the pore scale. The breakage criterium derived includes analytical expressions for tensile and shear stress maxima along with two geometric diagrams which allow determining the breaking stress. This leads to an explicit formula for the breakage flow velocity. Its upscaling yields a mathematical model for fines detachment by breakage, expressed in the form of the maximum retained concentration of attached fines versus flow velocity – maximum retention function (MRF) for breakage. We performed corefloods with piecewise constant increasing flow rates, measuring breakthrough concentration and pressure drop across the core. The behaviour of the measured data is consistent with two-population colloidal transport, attributed to detrital and authigenic fines migration. Indeed, the laboratory data show high match with the analytical model for two-population colloidal transport, which validates the proposed mathematical model for fines detachment by breakage.

## 1. Introduction

Dislodgement of natural reservoir fines from rock surfaces, induced by viscous flow in porous media, with the following migration and capture by the rock is essential in numerous natural and industrial processes. These include well fracturing, production of coal bed methane, water and polymer injection in aquifers and oilfields, heavy oil production, underground storage of $CO_2$ in aquifers and depleted oil and gas fields, fresh and hot water storage in aquifers, radioactive nuclear waste, and enhanced geothermal projects.[1-6] Usually, the migrating fines are clays (kaolinite, illite, chlorite), silica particles, or coals.[7-9] Fig. 1 shows SEM images of potentially migrating fines attached to rock surface. Figs. 1a and 1c show *authigenic* particles that naturally grow on rock surfaces during geological times, while Figs.1b and 1d show *detrital* particles attached to rock surfaces by electrostatic forces. Viscous flows in porous media induce drag and lift exerting on attached particles, which can result in their detachment. Authigenic particles are dislodged by stresses that initiate breakage of the particle-substrate bond, while detrital fines are detached by overcoming the electrostatic particle-substrate attraction. [10]

The detachment schematic at the pore scale is presented in Fig. 2a, where the detachment of authigenic and detrital particles occurs at the lower and upper parts of pore throat, respectively. The mobilisation of fines yields their straining in thin pore throats which consequently alters the fluid flow. The attached fines coat the rock surface, so their dislodging causes low-to-moderate permeability increase, while plugging the flow paths yields significant decline of permeability. [5, 11] The consequent decrease of well productivity and injectivity motivates significant efforts in studying migration of natural reservoir fines in porous media. [8, 12-14] Indeed, currently this topic is well developed for detrital fines. [12-14]

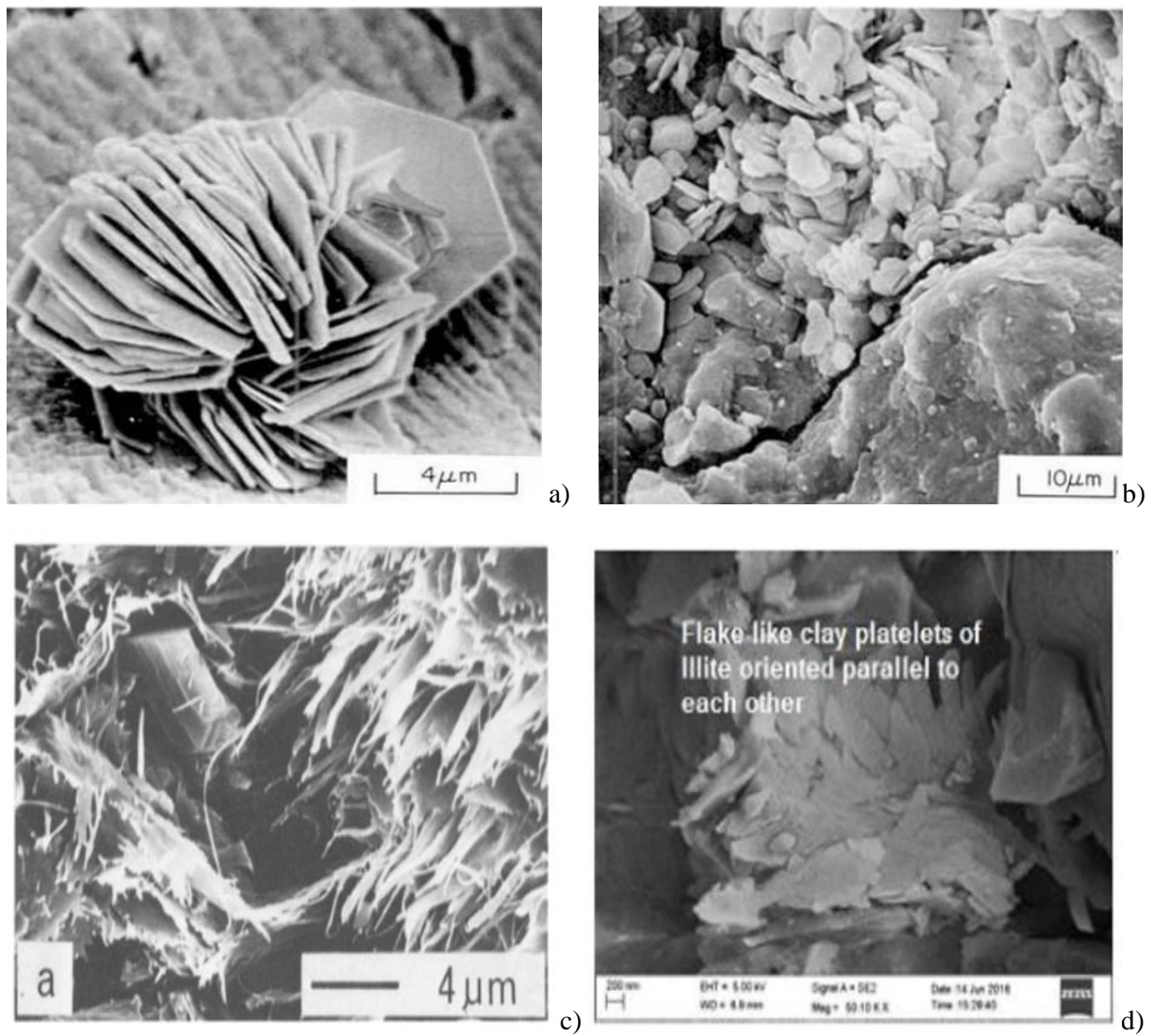

Figure 1: SEM photos of clay particles attached to the grains of sandstone rocks: a) authigenic kaolinite, [15] b) detrital kaolinite, [15] c) authigenic illite, [16] d) detrital illite[17]

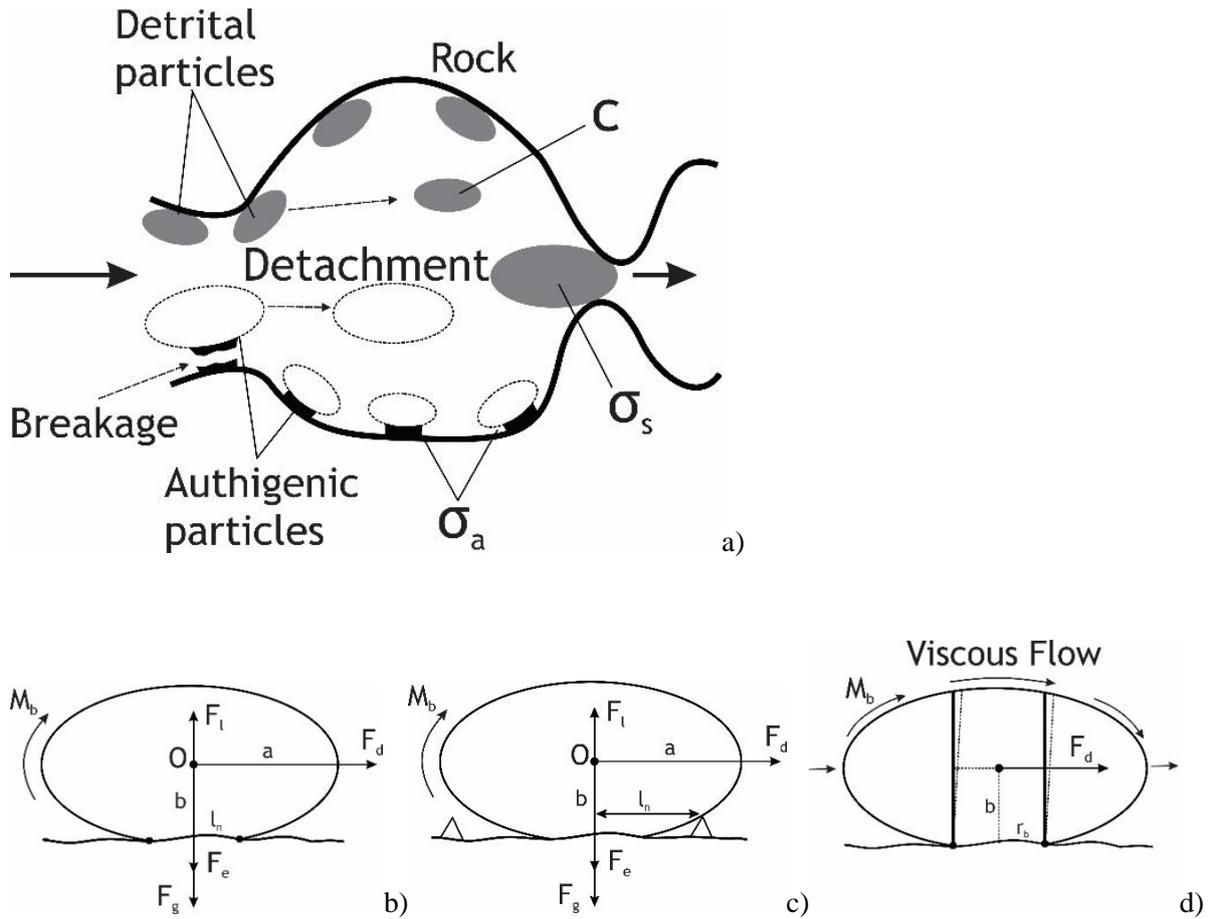

Figure 2. Detachment for detrital and authigenic clay particles: a) schematic for detachment at the pore scale; b) force (torque) balance at attached detrital fine with lever arm due to particle deformation; c) lever arm at attached detrital fine due to rock surface asperity; d) representation of attached authigenic particle by deformable beam[18]

Current mathematical and lab modelling for *detachment of detrital fines* is based on mechanical equilibrium of a particle situated on the solid substrate. [13, 19-21] Detrital fines detachment by drag against electrostatic forces is shown at the upper part of the entrance throat in Fig. 2a. Figs. 2b and 2c show drag, electrostatic, lift, and gravity forces exerting on an isolated fine particle. At the moment of dislodging, a particle rotates around a contact point on the rock surface. The corresponding lever arm is determined by either mutual particle-rock deformation, like in Fig. 2b, or by the size of the rock surface asperity, like in Fig. 2c. The attaching electrostatic force is described by DLVO theory. [21] The mathematical model for fines dislodging is either a linear-kinetics equation for detachment rate, [12, 14] or a function of retained concentration of attached particles versus velocity that is derived from mechanical equilibrium. [22, 23] Both models close the system of governing equations for colloidal-suspension-nano transport in porous media. In this work, to upscale the detachment model from pore to rock scale, we use the approach of maximum retention function (MRF). Whereas much is currently known about the underlying mechanisms governing the flow-induced detachment of detrital fines, a critical gap exists due to the absence of geomechanics-flow breakage criteria for authigenic fines.

*Detachment of authigenic fines* during flow in rocks, as it is shown in Figs. 2d, 3d, and 3h, occurs by breakage. Particle dislodgement by breakage under viscous flows in porous media has been observed during sand production, [24] well acidizing, [25, 26] cement dissolution in sandstones, [27, 28] carbonate rock dissolution in water, [29] and illite breakage under flow during hydraulic fracturing. [10] Guo et al. 2016 observed coal fines detachment by breakage during coreflood under piecewise-constant increasing velocity. [30] Wang et al. 2020 observed produced calcite particles from broken bonds with grains during waterflood tests. [26] Using SEM images, the above works clearly distinguish between detachment of detrital fines against electrostatic attraction and breakage of authigenic particles.

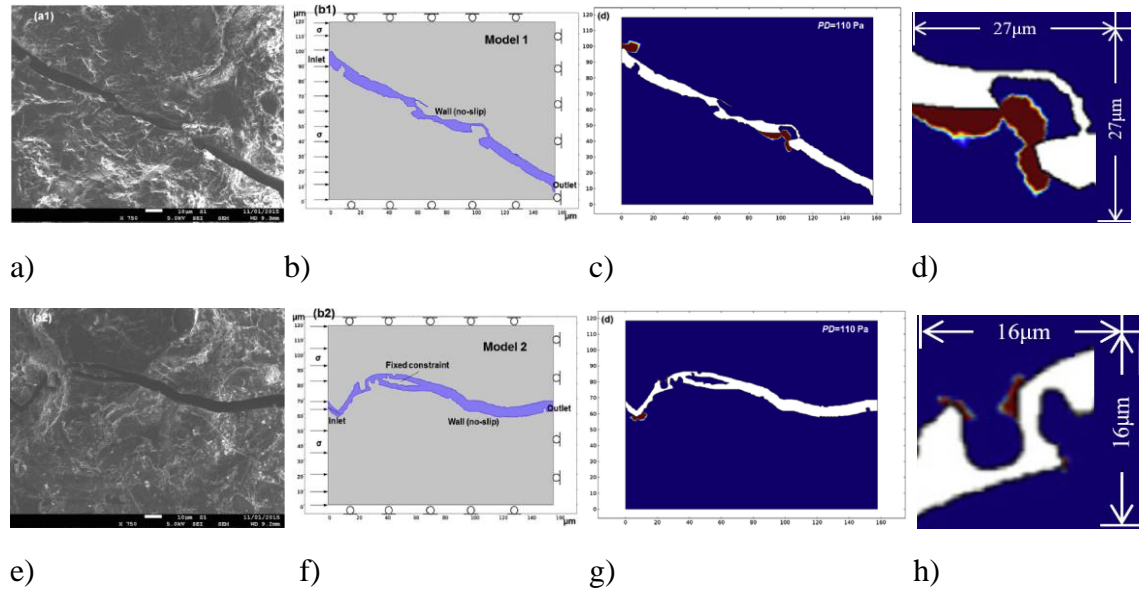

a)     b)     c)     d)

e)     f)     g)     h)

Figure 3: SEM photos of potential fine generation at different coal cleats (a and e), cleat and asperity geometries used in numerical simulations (b and f), failure zones (in red) for each case as a result of numerical simulation (c and g), and zoom for failure zones (d and h) [31]

The micro-scale numerical fines-detachment models couple flow and stress equations. [31, 32] SEM images (Figs. 3a, 3e) allow defining the channel model geometries (Figs. 3b, 3f), where the boundary conditions on the liquid-solid interfaces are posed and setting the detailed coupled numerical model for flow in porous channels and induced stresses in the rock (Figs. 3c, 3g). The failure zones are calculated from the stress field using various failure criteria (Figs. 3d, 3h), predicting particle detachment at the pore scale (Figs. 3d, 3h). The analytical model for stresses, induced by the external load, based on beam theory[33] was used to predict rock failure in sea-cliffs and sandstone canyon cliffs, [34, 35] and between grains consolidated by cement. [18, 36]

Micro scale modelling of particle detachment by breakage strongly depends on the rheological behaviour of particles bonded with the rock surface, the corresponding breakage criteria, and critical stress conditions. Laboratory pore-scale bending tests for single particles bonded to solid substrate have been performed for kerogen-rich shales, and brittle behaviour has been observed. [37] Other tests related to non-mineral-rock materials – polymer-clay, [38] glass, [39] and silicon[40-42]– also exhibited brittle behaviour, while the tests with nickel showed ductile behaviour. [43] Geo-mechanical tests with partly water-saturated kaolinite powder detect brittle behaviour, [44-49] while for high saturations of water, the powder becomes ductile. [44, 50, 51] Mixing kaolinite soil with more than 1% of cement changes ductile behaviour of the stress-strain diagram into brittle. [52] Quartz powder and its rich mixtures with kaolinite exhibit brittle behaviour. [53] Carbonate powders, and their rich mixtures with quartz in low water saturations show typical brittle stress-strain diagrams. [53, 54] Moreover, sand particles bonded with calcium carbonate powder under tensile and shear tests show shape decline after failure which is an indication of brittle behaviour. [54]

Other laboratory studies also encountered either brittle or ductile failure in reservoir rocks, in particular in kaolinite-rich rocks, [55, 56] illite-rich shales, [57-59] chlorite-rich black shale rocks, [60] quartz-rich sandstone, [61, 62] cement mortar rock, [63] and coal rocks. [64-66]

Molecular-dynamic simulation of rheological kaolinite behaviour for hydrated and defected kaolinite crystals with the typical length of 100 Angstrom shows brittle properties under stress loading at the tension case with parallel and perpendicular to layering. [67-69] Zhang et al. 2021 simulate both tension and compaction loads; tension stress-strain diagrams have brittle type for load parallel and perpendicular to layering, while those for compression are brittle for parallel load and are ductile for perpendicular load. [70]

The above experimental studies highlight the prevalence of mechanical failure of colloidal particles in porous media. The difference in detachment criteria for authigenic and detrital particles is important for understanding and modelling fines migration. However, this distinction hasn't been used in the analysis of coreflooding or field production data. Despite these phenomena being widely spread, a pore-scale mathematical model and its upscaling to the rock scale transport are not available.

The present paper fills the gap. This contribution integrates CFD-based modelling of viscous fluid – attached particle interaction, 3D elastic beam theory, and strength failure criteria, yielding an explicit expression for breakage detachment conditions of authigenic fines. It was found that stress maxima are reached at either the middle or boundary of the beam base. Introduction of tensile-stress and shear-tensile diagrams allows determining which stress causes the particle failure. Formulae for breakage flow velocities have been derived for all cases of particle breakage by different stresses. The expressions for breakage velocity allow determining the maximum retention concentration versus velocity (MRF), which is a mathematical model for fines mobilisation by breakage at the rock (laboratory cores and reservoirs) scale. The laboratory test undertaken comprises coreflooding with 7 rates taken in increasing order while measuring particle breakthrough concentrations and pressure drop across the core. High match between the model and the experimental data from this test, and also from 4 tests taken from the literature, validate the mathematical model for particle detachment by breakage.

**2. Microscale model for particle detachment from rock surface by breakage**

This section integrates Timoshenko's beam theory with creeping viscous flow around attached particles. This includes assumptions of the model (section 2.1), CFD-based expressions for drag and torque for the particles with different geometries (section 2.2), and derivations for tensile and shear stress distributions over the beam base (section 2.3).

*2.1. Assumptions of the particle breakage model*

The breakage detachment model for a single particle is based on Navier-Stokes equations for viscous flow around the attached particle with resulting drag force and moment exerting on the particle, elastic beam theory, [33] and the rock failure criteria by tensile or shear strength. [71-73] The main detaching force is drag (Figs. 2b, 2c). We assume small deformation for solid mineral particles and negligible effect of particle deformation on the drag force and moment.

Fig. 2d shows 3D cantilever beam for spheroidal particle; the undeformed vertical configuration is exhibited by continuous lines; the end loading by drag displaces material points to the deformed shape shown by dashed curves. The stresses in the particle outside the beam are lower than those inside, which justifies the beam approximation of an irregularly shaped particle for deformation modelling. The assumption that the particle volume around the beam stem has negligible impact on the stress maxima over the particle-substrate contact area has been used by Robinson et al. 1970, Young and Ashford 2008, Obermayer et al. 2013, Wagner et al. 2016, and Chen et al. 2022. [18, 34, 35, 74, 75]

Because drag is applied to the centre of mass of the particle, the beam connects the base to the centre, and the drag acts as an external load on the top cross-section of the beam. We assume that the particle shape is spheroidal, and the contact area is circular. Cylindrical shaped particles are considered too. We also assume homogenous and linear-elastic particle matter.

It is assumed that planar sections perpendicular to the neutral axis before deformations remain planar, but not necessarily perpendicular to the neutral axis after deformation, i.e., shear deformations cannot be ignored. This is particularly important for "short" kaolinite and chlorite clay particles that represent the most widely spread fines in natural reservoirs. Therefore, stress modelling for fines breakage is based on Timoshenko's rather than Bernoulli-Euler beam theory. [75]

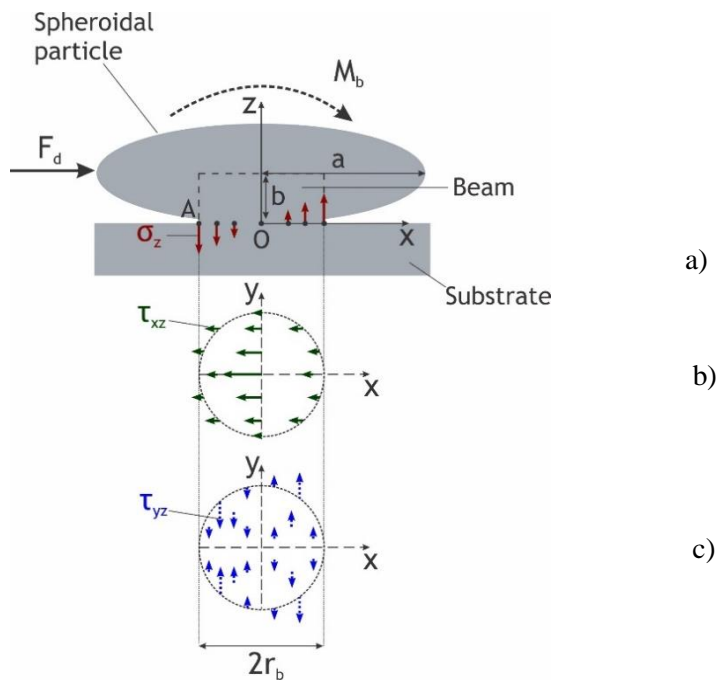

Figure 4: Schematic of equivalent beam for attached spheroidal particle: a) loading force and moment exerting from viscous flow; b) shear in plane parallel/perpendicular by Timoshenko's solution; c) shear in plane parallel/perpendicular by Timoshenko's solution

The present work assumes that the maximum stress in the particle-substrate contact area due to drag is determined by the deformation of the cylindrical beam with the base equal to the particle-substrate contact area. This assumption was adopted from Obermayr et al. 2013. [18] Therefore, the failure criteria for the attached particle is determined from the cylindric beam deformation from Timoshenko's solution. The same assumption has already been used by Robinson, 1970, and Young and Ashford, 2008. [34, 35]

Figs. 2b, 2c show the spheroidal particle, drag exerting from the moving viscous fluid, and the induced moment. Under slow creeping flows in porous media, lift is negligibly small if compared with drag . Gravity can also be ignored. Fig. 4a shows the normal stress $\sigma_z$ versus the horizontal x-coordinate at the beam base z=0. The advancing point of the particle is in extension as a result of the drag, and the receding point is in compression. Figs. 4b and 4c exhibit the distributions of shear stresses ($\tau_{xz}$ and $\tau_{yz}$) over the beam cross section. Stress $\tau_{xz}$ is maximum at central point of the base, while stress $\tau_{yz}$ is zero at this point. Timoshenko's beam model assumes that normal stress over a cross section is distributed in the same manner as in the case of pure bending. The remaining three stress components i.e. $\sigma_x$, $\sigma_y$, and $\tau_{xy}$ are zero.[33] The expressions for all stresses of the beam theory are presented in Appendix A.

The particle exhibits brittle behaviour with breakage. The breakage occurs instantly according to maximum stress criteria, i.e., if either tensile or shear stress reaches strength (maximum) values.[71, 72]

*2.2. Drag force and moment*

Consider Couette flow of viscous fluid over a plane substrate and around the attached particle (Figs. 2b, 2c, 2d). Drag $F_d$ and its moment $M_b$ for spherical, oblate spheroidal, and cylindrical particles are extensions of the Stokes formula, which is valid for spherical particles[76]:

$$F_d = 6\pi\mu_f r_s V f_d(\alpha_s) \tag{1}$$

$$M_b = F_d b f_M(\alpha_s) = 6\pi\mu_f r_s V f_d(\alpha_s) b f_M(\alpha_s) \tag{2}$$

where, $\mu_f$ is the fluid viscosity, $r_s$ is the particle radius, $V$ is the interstitial fluid velocity, $f_d$ is the shape factor for drag force, $b$ is the lever arm for drag force, and $f_b$ is the moment shape factor. Interstitial fluid velocity $V$ is expressed via Darcy's velocity $U$ as

$$U = \phi V \tag{3}$$

where $\phi$ is the porosity.

For a non-spherical particle, the effective radius, $r_s$ is determined based on the equality of the volume of the desired shape and a sphere with radius $r_s$.

For a cylindrical particle, the effective radius and aspect ratio, $\alpha_s$ are defined as

$$2\pi a_c^2 b_c = \frac{4}{3}\pi r_s^3, \quad r_s = b_c\left(\frac{2}{3}\alpha_s^2\right)^{-1/3}, \quad \alpha_s = b_c/a_c \tag{4}$$

where $a_c$ and $b_c$ are the cylinder base radius and height, respectively.

The effective radius $r_s$ and aspect ratio for a spheroidal particle are defined similarly as:

$$\pi a^2 b = \frac{4}{3}\pi r_s^3, \quad r_s = b\alpha_s^{-2/3}, \quad \alpha_s = b/a \tag{5}$$

where $a$ and $b$ are semi-major and semi-minor axes of spheroid, respectively, and $\alpha_s$ is the aspect ratio.

Following Ting et al. 2021, we calculate the shape factors for drag force and moment using CFD package ANSYS/CFX.[77] The calculations are performed for long thin cylinders, which approximate illite clay particles (Figs. 1c, 1d). Figs. 2a, 2b, 2c show the schematic for fluid flow around the attached particle. Drag is calculated from the solution of the Navier-Stokes equations for viscous flow with no-slip conditions at the solid-fluid particle and substrate surfaces. The shape factors for drag force and moment, $f_D$ and $f_M$, respectively, are calculated from the numerical solution using Eqs. (1) and (2).

For long thin cylinders, which correspond to aspect ratio $\alpha_s>1$, the correlations for drag force and moment factors, based on multiple runs of the CFD package, are:

$$f_d = \left(0.9014\alpha_s^2 + 1.599\alpha_s + 2.265\right)\left(\alpha_s + 1.752\right)^{-1} \tag{6}$$

$$f_M = \left(0.0002161\alpha_s^3 + 1.34\alpha_s^2 + 44.18\alpha_s + 21.27\right)\left(\alpha_s^2 + 31.34\alpha_s + 10.38\right)^{-1} \tag{7}$$

respectively. We will be using these correlations further in the text to calculate maximum stresses and predict the particle-substrate bond breakage. The expressions for drag and moment factors for spheroidal and thin-cylinder particles, which model kaolinite, chlorite, and silica particles, are available from Ting et al. 2021.[77]

*2.3. Stress distributions at the base of the beam*

Substitution of beam stress equations (A1-A3) into the expressions for principal stresses (A5), yields the equations for maximum tensile and shear stresses at the beam base:

$$\sigma_3 = \frac{1}{2}\frac{F_d}{I}r_b^2\left(\frac{bf_M}{r_b}\frac{x}{r_b} - \sqrt{\left(\frac{bf_M}{r_b}\right)^2\left(\frac{x}{r_b}\right)^2 + 4\left(\left(\frac{(3+2\upsilon)}{8(1+\upsilon)}\left(1-\left(\frac{x}{r_b}\right)^2 - \frac{(1-2\upsilon)}{(3+2\upsilon)}\left(\frac{y}{r_b}\right)^2\right)\right)^2 + \left(-\frac{(1+2\upsilon)}{4(1+\upsilon)}\frac{x}{r_b}\frac{y}{r_b}\right)^2\right)}\right) \tag{8}$$

$$\frac{\sigma_1-\sigma_3}{2} = \frac{1}{2}\frac{F_d}{I}r_b^2\left(\sqrt{\left(\frac{bf_M}{r_b}\right)^2\left(\frac{x}{r_b}\right)^2 + 4\left(\left(\frac{(3+2\upsilon)}{8(1+\upsilon)}\left(1-\left(\frac{x}{r_b}\right)^2 - \frac{(1-2\upsilon)}{(3+2\upsilon)}\left(\frac{y}{r_b}\right)^2\right)\right)^2 + \left(-\frac{(1+2\upsilon)}{4(1+\upsilon)}\frac{x}{r_b}\frac{y}{r_b}\right)^2\right)}\right) \tag{9}$$

Axes are shown in Fig. 4.

Eqs (8) and (9) for normalised stresses can be transformed into the following dimensionless form

$$\frac{\sigma_3}{T_0} = \frac{1}{\kappa}\left(X - \sqrt{X^2 + \chi\left(1-X^2 - \frac{(1-2\upsilon)}{(3+2\upsilon)}Y^2\right)^2 + \chi\frac{4(1+2\upsilon)^2}{(3+2\upsilon)^2}(XY)^2}\right) \tag{10}$$

$$\frac{\sigma_1-\sigma_3}{2S_0} = \frac{\eta}{\kappa}\sqrt{X^2 + \chi\left(1-X^2 - \frac{(1-2\upsilon)}{(3+2\upsilon)}Y^2\right)^2 + \chi\frac{4(1+2\upsilon)^2}{(3+2\upsilon)^2}(XY)^2} \tag{11}$$

where $T_0$ and $S_0$ are the tensile and shear strengths, respectively. Normalised stress expressions (10, 11) contain three dimensionless groups reflecting the interaction between the creeping flow around an attached particle and the induced elastic deformation of the particle – dimensionless numbers $\kappa$, $\chi$, and $\eta$:

$$\kappa = \frac{2T_0}{F_d} \frac{I}{r_b b f_M} = \frac{2T_0}{F_d} \frac{I}{\delta \alpha_s f_M}, \quad \chi = \left[\frac{r_b}{bf_M} \frac{3+2\upsilon}{4(1+\upsilon)}\right]^2 = \left[\frac{\delta}{\alpha_s f_M} \frac{3+2\upsilon}{4(1+\upsilon)}\right]^2, \quad \eta = \frac{T_0}{S_0}, \quad \delta = \frac{r_b}{a}, \quad X = \frac{x}{r_b}, \quad Y = \frac{y}{r_b} \quad (12)$$

Dimensionless group $\kappa$ is proportional to the ratio between tensile strength and fluid pressure caused by the drag exerting on the particle cross section and is called the *strength-drag number*. Dimensionless number $\kappa$ also depends on bond ratio $\delta$, aspect ratio $\alpha_s$, and moment of inertia $I$. Dimensionless group $\chi$ depends on geometric parameters, namely the bond ratio $\delta$ and aspect ratio $\alpha_s$, and on the Poisson ratio $\upsilon$ and is called, therefore, the *shape-Poisson number*. The *strength number* $\eta$ is the ratio between tensile and shear strengths.

Common interval of parameters for rock minerals are: aspect ratio $\alpha_s$ varies from 0.03 to 1.0 for spheroids and flat cylinders and from 1 to 100 for long cylinders, bond ratio $\delta$ - from $10^{-3}$ to 1.0, and Poisson's ratio $\upsilon$ − from zero to 0.5. Eq. (12) along with formulae for drag and moment shape factors results in values of the shape-Poisson number $\chi$ varying from $2.4 \times 10^{-7}$ to 4.8 for spheroids, from $2.4 \times 10^{-5}$ to 0.3 for long cylinders, and from 0.18 to 135 for flat cylinders.

## 3. Derivation of maxima for tensile and shear stresses

Here we transform a graphical technique to determine the stress that meets the strength failure criteria. It includes the derivation of stress maxima at the beam base (section 3.1), calculation of tensile stress maxima over the beam base middle and boundary (section 3.2) and their comparison using the tensile stress diagram (section 3.3), calculation of shear stress maxima over the beam base middle and boundary (section 3.4) definition of the failing stress using the tensile-shear diagram (section 3.5).

The failure criteria used in this work correspond to reaching the strength values by maximum tensile and shear stresses.[71,73] The corresponding expressions for tensile and shear failures are:

$$-\sigma_3 \geq T_0, \quad \frac{-\sigma_3}{T_0} \geq 1 \quad (13)$$

$$\frac{\sigma_1 - \sigma_3}{2} \geq S_0, \quad \frac{\sigma_1 - \sigma_3}{2S_0} \geq 1, \quad (14)$$

respectively. Here $T_0$ is the tensile strength, and $S_0$ is the shear strength.

To determine which stress causes failure, the following maxima must be compared

$$\max_{X^2+Y^2 \leq 1} \frac{-\sigma_3(X,Y)}{T_0}, \quad \max_{X^2+Y^2 \leq 1} \frac{\sigma_1(X,Y) - \sigma_3(X,Y)}{2S_0} \quad (15)$$

To apply failure criteria (13, 14) to the expressions for normalised tensile and shear stresses at the base of the beam, in the next section we calculate the maxima of those stress functions over the area $X^2+Y^2 \leq 1$.

*3.1. Stress maxima at the base of the beam*

Consider maxima of the tensile and shear stresses given by Eqs. (10, 11). If the maxima points $(X_m, Y_m)$ are located inside the base circle, $X_m^2 + Y_m^2 < 1$, partial derivatives of both expressions (10) and (11) over $Y$ must be zero. Both expressions depend on $Y^2$, so the expressions for first partial derivatives in $Y$ contain $Y$ as a multiplier and is zero at $Y=0$. It is possible to show that the multiplier inside the unitary circle is positive, and that second partial derivatives in $Y$ of both expressions (10) and (11) are negative at $Y=0$. Therefore, all maxima inside the base circle $X_m^2 + Y_m^2 < 1$ are reached along the middle of the base, i.e., axis $Y=0$. Otherwise, tensile or shear stresses reaches maxima at the beam base over the boundary $X_m^2 + Y_m^2 = 1$.

Expressions for normalised tensile and shear stresses versus $X$ in the *beam middle* $Y=0$ are obtained from Eqs. (10, 11):

$$\frac{\sigma_3}{T_0} = T^0(X, \chi) = \frac{1}{\kappa}\left(X - \sqrt{X^2 + \chi(1-X^2)^2}\right) \tag{16}$$

$$\frac{\sigma_1 - \sigma_3}{2S_0} = S^0(X, \chi) = \frac{\eta}{\kappa}\sqrt{X^2 + \chi(1-X^2)^2} \tag{17}$$

Expressions for normalised tensile and shear stresses versus $X$ at the *cylinder boundary* are obtained from Eqs. (10, 11) by substituting $Y^2 = 1-X^2$:

$$\frac{\sigma_3}{T_0} = T^1(X, \chi, \nu) = \frac{1}{\kappa}\left(X - \sqrt{X^2 + \chi\frac{4(1+2\upsilon)^2}{(3+2\upsilon)^2}(1-X^2)}\right) \tag{18}$$

$$\frac{\sigma_1 - \sigma_3}{2S_0} = S^1(X, \chi, \nu) = \frac{\eta}{\kappa}\sqrt{X^2 + \chi\frac{4(1+2\upsilon)^2}{(3+2\upsilon)^2}(1-X^2)} \tag{19}$$

*3.2. Maxima of tensile stresses in the beam middle and boundary*

In order to apply failure criteria (13, 14) to expressions for normalised tensile and shear stresses at the cylinder middle and at boundary, let us calculate maxima of the four stress functions (16-19) over the closed $X$-interval [-1,1]. Fortunately, for all 4 cases, the maximum points $X_m$ and the corresponding values of normalised stresses can be found explicitly. In the middle of the base, maxima of the two functions (16) and (16) inside the base circle (-1,1) are determined by conditions of zero first derivative in $X$ at some point $X=X_m$, and negative second derivative in the same point. Then the obtained maxima are compared with stresses at the boundary $X=-1$ and $X=1$. The same procedure is applied for stress functions (18, 19) on the beam base boundary. Afterward, the detachment regime and the detachment point $X_m$ are determined by comparison of the 4 normalised tensile and shear stresses at the boundaries $X=-1$ and $X=1$, and the open interval between them.

From now on, we call dimensionless stress the product of strength-drag number $\kappa$ and normalised stress, defined in Eqs. (16-19).

The profiles for dimensionless tensile stress $\kappa T_m^0(X)$ in the middle of the cylinder base for four values of the shape-Poisson number $\chi=1, 2, \chi=\chi_1$, and $\chi=4.5$ are presented in Fig. 5a. Here $\chi=\chi_1$ is the value where $\kappa T_m^0$ at the boundary $X=-1$ is equal to maximum inside the interval. As $\chi$ tends to zero, the plot of $\kappa T_m^0(X)$ tends to two straight lines corresponding to two values of square root in expressions (34, 35). At some $\chi$ there does appear a maximum inside the open interval (-1,1), which remains below $\kappa T^0_m(-1,\chi)=2$, reached in the advanced point $X_m=-1$. Inequality $\kappa T^0_m < 2$ remains fulfilled for $\chi < \chi_1$. The threshold value $\chi_1$ and corresponding maximum point $X=X_m$ are determined from system of two transcendental equations and one inequality:

$$\kappa T^0(X_m, \chi_1) = 2, \quad \frac{\partial \kappa T^0}{\partial X}(X_m, \chi_1) = 0, \quad \frac{\partial^2 \kappa T^0}{\partial X^2}(X_m, \chi_1) < 0 \qquad (20)$$

The solution of (36) is unique; the roots are found numerically: $\chi_1 = 3.38$, $X_{m1} = -0.33$.

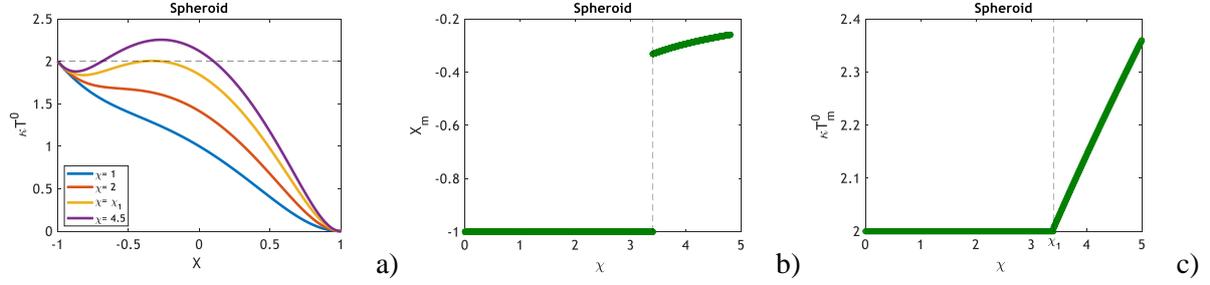

Fig. 5. Maximum tensile stress at Y=0: a) dimensionless tensile stress in the middle versus $\chi$ for 3 different values of aspect-Poisson number $\chi$; b) position of maximum tensile stress $X_m$ at Y=0; c) maximum value of tensile stress at Y=0 versus $\chi$

For $\chi > \chi_1$, the maximum is reached inside the open interval $-1 < X_m < 1$ (violet curve in Fig. 5a). The expression for maximum normalised tensile stress at Y=0 is

$$T_m^0(\chi) = \frac{1}{\kappa}\begin{cases} 2, & \chi \leq \chi_1 \\ -X_m(\chi) + \sqrt{X_m^2 + \chi(1-X_m^2)^2}, & \chi > \chi_1 \end{cases}; \quad X_m = \begin{cases} -1, & \chi \leq \chi_1 \\ -\sqrt{-\frac{\sqrt{(4\chi-1)(4\chi-9)} - 4\chi + 3}{8\chi}}, & \chi > \chi_1 \end{cases} \qquad (21)$$

Substituting $X_m$ into $T^0_m(\chi)$ in Eq. (21), we obtain

$$T_m^0(\chi) = \frac{1}{\kappa}\begin{cases} 2, & \chi \leq \chi_1 \\ \sqrt{-\frac{\sqrt{(4\chi-1)(4\chi-9)} - 4\chi + 3}{8\chi}} + \sqrt{-\frac{\sqrt{(4\chi-1)(4\chi-9)} - 4\chi + 3}{8\chi} + \chi\left(1 + \frac{\sqrt{(4\chi-1)(4\chi-9)} - 4\chi + 3}{8\chi}\right)^2}, & \chi > \chi_1 \end{cases} \qquad (22)$$

Figs. 5b and 5c show the plot of maximum point $X_m$ and maximum tensile stress in the middle $T^0_m(\chi)$ versus the aspect-Poisson ratio. For $\chi$ lower than $\chi_1$, the maximum is reached in the advance point $X_m=-1$. For higher $\chi$-values, the point $X_m$ jumps to $X_{m1}$ and then continuously moves to the right towards the origin.

Eq. (34) for dimensionless tensile strength on the base boundary shows that $T^1_m$ depends on the dimensionless group $\xi$ which is proportional to $\chi$. The proportionality coefficient depends on Poisson ratio:

$$\xi = \chi \frac{4(1+2\upsilon)^2}{(3+2\upsilon)^2} \qquad (23)$$

Fig. 6a shows the profiles for dimensionless tensile stress on the beam boundary for different values of $\xi$. At $\xi < 2$, $\kappa T^1_m(X,\chi)$ monotonically decreases versus $X$, so the maximum is reached at the point $X_m=-1$. At $\xi=2$, the slope of profile at $X=-1$ reaches zero, and at $\xi>2$ a maximum is reached inside the interval at $X>-1$. The expressions for maximum $T^1_m(\xi)$ and $X_m$ are

$$T_m^1(\chi,\nu) = \frac{1}{\kappa}\begin{cases} 2, & \xi \leq 2 \\ \xi(\xi-1)^{-0.5}, & \xi > 2 \end{cases}; \quad X_m = \begin{cases} -1, & \xi \leq 2 \\ -(\xi-1)^{-0.5}, & \xi > 2 \end{cases} \quad (24)$$

Figs. 6b and 6c show the plots for maximum point and value vs $\chi$ at three Poisson ratios. For low $\chi$, determined by condition $\xi<2$, the maximum is reached at the advance point $X=-1$ and is equal to 2. For higher $\chi$, the maximum point continuously moves right from the advance point, and the maximum value monotonically increases from 2.

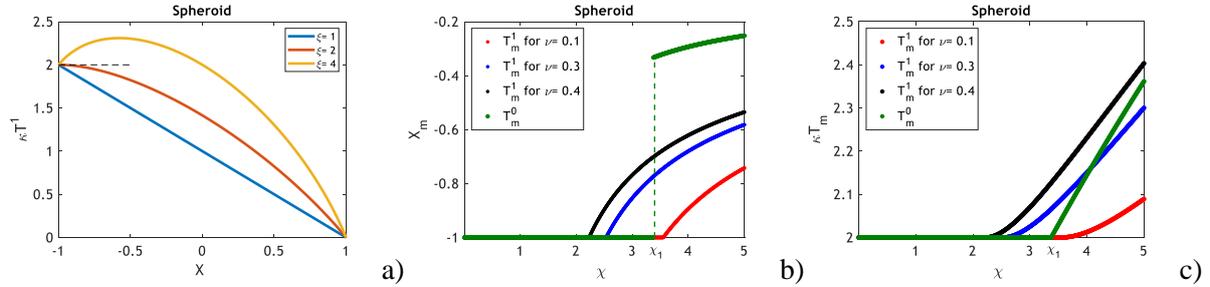

Fig. 6. Maximum tensile stress at beam boundary: a) dimensionless tensile stress on the beam boundary versus $\chi$ for 3 different values of $\nu$; b) position of maximum tensile stress $X_m$ at the beam boundary; c) maximum value of tensile stress over the boundary versus $\chi$

*3.3. Comparison of maximum tensile stresses in the beam middle and boundary*
Consider the stress equality, separating the domains in ($\chi,\upsilon$) plane where either of the stresses dominate

$$T_m^0(\chi) = T_m^1(\chi,\nu) \quad (25)$$

The black curve in Fig. 7a corresponds to $\xi=2$; the equation for the black curve follows from the expression (23):

$$\chi = \frac{(3+2\upsilon)^2}{2(1+2\upsilon)^2}, \quad \upsilon = \left(1-\sqrt{\frac{\xi}{4\chi}}\right)^{-1} - \frac{3}{2}. \quad (26)$$

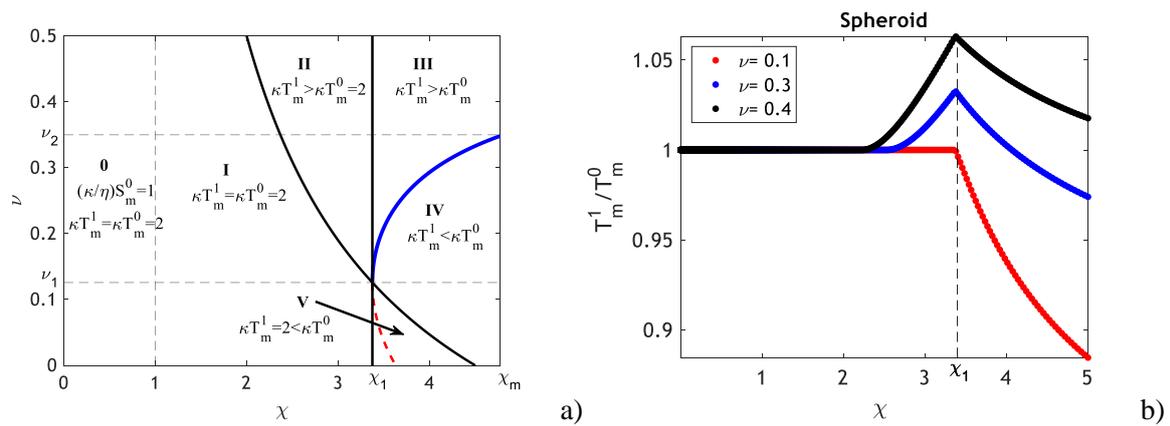

Fig. 7. Comparison between dimensionless tensile stresses in the middle and on the beam boundary: a) determination of the critical Poisson ratio whether the tensile stresses are equal; b) the ratio between tensile stresses on the boundary and in the middle

The straight line $\chi=\chi_1$ and curve (26) for $\xi=2$ separate plane $(\chi,\upsilon)$ into 4 domains that correspond to different expressions (22) and (24) for maximum tensile stresses (Fig. 7a). Three lines cross in point with $\chi=\chi_1$ and $\upsilon=\upsilon_1=0.125$.

Here the variables are limited by maximum $\chi$-values for natural minerals: $\chi<5$.

Altogether 5 different domains that defines maxima between $T^0_m$ and $T^1_m$ can be distinguished:

I. In the case $\chi<\chi_1$ and $\xi<2$, both tensile stresses are equal to 2.
II. In domain $\chi<\chi_1$ and $\xi>2$, $T^0_m=2$. Equation (25) for the boundary between regions I and II becomes: $T^1_m(\chi,\upsilon)=2$.

This equation has only one root: $\xi=2$. For $\xi>2$, $T^1_m$ monotonically increases. Therefore, in this domain $T^1_m>T^0_m$.

Consider the case where $\chi>\chi_1$ and $\xi>2$. In this region, both $\kappa T^0_m$ and $\kappa T^1_m$ are greater than 2. The region can be divided into two regions in which each stress is greater. Substituting the second line of Eq. (24) into equality (25) and expressing $\xi$ yields

$$\xi=\frac{\left(\kappa T^0_m(\chi)\right)^2\pm\kappa T^0_m(\chi)\sqrt{\left(\kappa T^0_m(\chi)\right)^2-4}}{2}; \chi>\chi_1 \qquad (27)$$

Critical Poisson's ratios where the tensile stresses in the middle and on the boundary are equal are determined by substitution of expression (23) for $\xi$ into Eq. (27) and solving for $\upsilon$:

$$\upsilon_m(\chi)=\frac{3\sqrt{\left(\kappa T^0_m(\chi)\right)^2\pm\kappa T^0_m(\chi)\sqrt{\left(\kappa T^0_m(\chi)\right)^2-4}}-\sqrt{8\chi}}{4\sqrt{2\chi}-2\sqrt{\left(\kappa T^0_m(\chi)\right)^2\pm\kappa T^0_m(\chi)\sqrt{\left(\kappa T^0_m(\chi)\right)^2-4}}} \qquad (28)$$

Explicit expression $\upsilon=\upsilon_m(\chi)$ is determined by equating the second lines of Eqs. (22) and (24) and is very cumbersome.

Fig. 7a shows the plot $\upsilon=\upsilon_m(\chi)$. It consists of two branches – the blue branch corresponds to the positive square root in Eq. (28), and the red branch to the negative root. Over the red branch, we have $\xi<2$, so only the blue branch belongs to the domain $\chi>\chi_1$ and $\xi>2$. This determines zones III and IV:

III. In the case where $\chi>\chi_1$ and $\upsilon>\upsilon_m(\chi)$, $T^1_m>T^0_m$.
IV. In the case where $\chi>\chi_1$ and $\upsilon<\upsilon_m(\chi)$, $\xi>2$, $T^1_m<T^0_m$.

The blue curve (45) crosses the line $\chi=\chi_1$ at the point with ordinate $\upsilon_1$.

V. In the case $\chi>\chi_1$ and $\xi<2$, $T^1_m=2$. As it follows from definition (20) of $\chi_1$, equation $T^0_m(\chi)=2$ has root $\chi=\chi_1$. No more roots exist for $\chi>\chi_1$. Therefore, in this domain $T^0_m>T^1_m=2$.

Five domains I, II…V in the so-called tensile stress diagram (Fig. 7a) determine where maximum tensile stress is higher – at the boundary of the beam base or in the middle. Their ratio versus shape-Poisson number is presented in Fig. 7b for three value of Poisson ratio: $\upsilon<\upsilon_1$, $\upsilon_1<\upsilon<\upsilon_2$, and $\upsilon>\upsilon_2$.

## 3.4. Maxima shear stresses in the beam middle and boundary

Fig. 8a shows profiles for shear stress in the middle of the beam base for different $\chi$. For $\chi$ tending to zero, the profile tends to two straight lines. At higher $\chi$, there does appear a maximum at $X_m=0$, which reaches unitary value at $\chi = \chi_2 = 1$. At $\chi < \chi_2$, the maximum remains in the advancing and receding points, for $\chi > \chi_2$ it moves to the origin. Expressions for stress maximum are obtained from Eq. (17):

$$S_m^0(\chi) = \frac{\eta}{\kappa}\begin{cases}1, & \chi<1\\ \sqrt{\chi}, & \chi>1\end{cases}; \quad X_m = \begin{cases}\pm 1, & \chi<1\\ 0, & \chi>1\end{cases}; \quad \eta = \frac{T_0}{S_0} \qquad (29)$$

Fig. 8b shows that the maximum points lie at the edges of the beam middle at $X_m=-1$ and $X_m=1$ for $\chi<\chi_2$, then moves to the centre point, $X=0$ for $\chi>\chi_2$. The maximum shear is equal to one for $\chi<\chi_2=1$, and monotonically increases from 1 for $\chi>\chi_2$ (Fig. 8c).

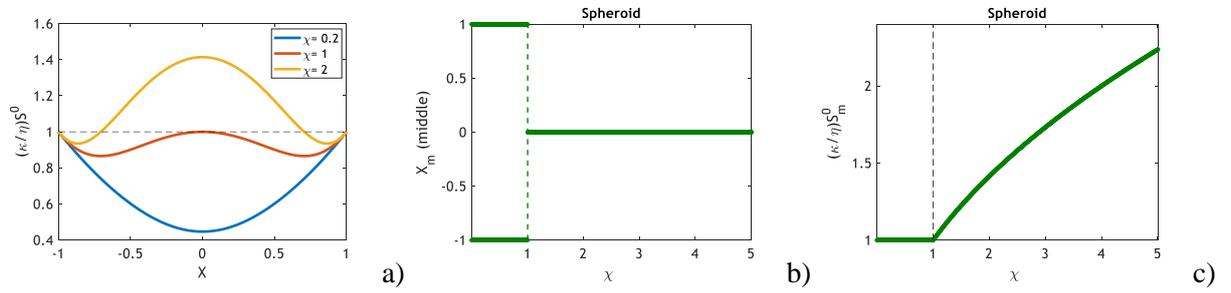

Fig. 8. Dimensionless maximum shear stress in the middle of the beam base: a) profiles for dimensionless stress in the middle of the beam base for different versus $\chi$; b) position of the maximum point $X_m$; c) maximum dimensionless stress versus $\chi$

The expression for maximum tensile stress along the boundary is obtained from Eq. (19):

$$S_m^1(\chi_1) = \frac{\eta}{\kappa}\begin{cases}\sqrt{\xi}, & \xi \geq 1\\ 1, & \xi<1\end{cases}; \quad X_m = \begin{cases}0, & \xi \geq 1\\ \pm 1, & \xi<1\end{cases} \qquad (30)$$

Corresponding profiles and maximum plots and presented in Fig. 9.

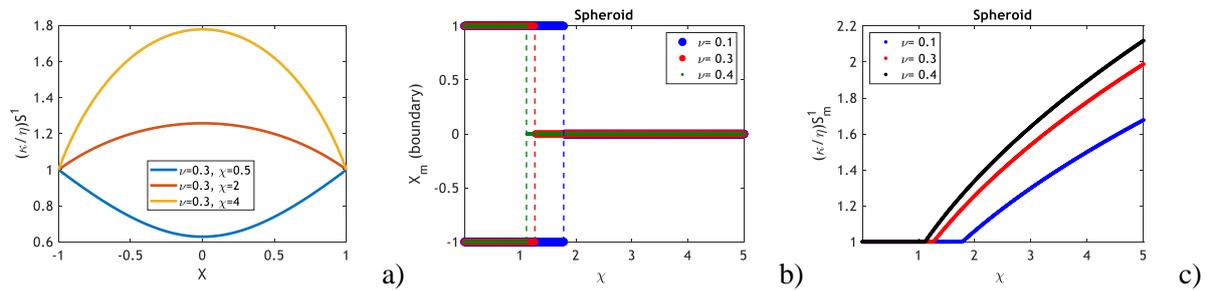

Fig. 9. Dimensionless maximum shear stress on the beam boundary: a) profiles for dimensionless stress for different $\nu$; b) position of the maximum point $X_m$; c) maximum dimensionless stress versus $\chi$

Fig. 10 shows the ratio $S^1_m/S^0_m$ versus $\chi$ for 3 values of Poisson ratio. For all parameter values, the ratio does not exceed one, i.e. $S^1_m < S^0_m$. Therefore, further in determining the breakage regime, shear at the base boundary is not considered.

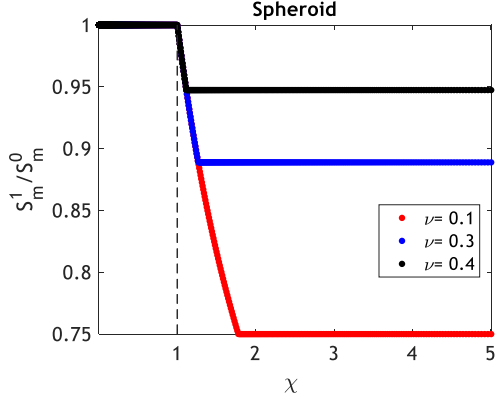

Fig. 10. Comparison between maximum shear stresses in the middle and on the boundary

Introduce region 0 in $(\chi,\upsilon)$-plane for $\chi<1$ (tensile-stress diagram in Fig. 7a). According to Eq. (29), here the dimensionless shear is equal to one, while both tensile stress maxima are equal to two.

### 3.5. Determination of breakage regime using the breakage function

This section compares maxima of tensile and shear stresses, Eqs. (22, 24, 29). The tensile-stress diagram in Fig. 11a shows which maximum tensile stress – in the beam middle or at the boundary – is higher. Shear stress in the middle is always higher than that at the boundary. Now consider breakage by tensile stress, where equality (13) is fulfilled at some point $(X_m,0)$ and normalised tensile stress reaches its maximum which equals one, while equality (14) is not fulfilled for any point $-1<X<1$. Consequently, the condition for breakage by tensile stress is

$$\frac{\max(-\sigma_3)}{T_0} \geq \frac{\max(\sigma_1 - \sigma_3)}{2S_0} \tag{31}$$

Define the breakage regime function, $g(\chi,\upsilon)$ as the ratio of the dimensionless tensile stress to the dimensionless shear stress. Comparing with Eq. (31), we arrive at:

$$g(\chi,\upsilon) = \frac{\kappa T_m(\chi,\upsilon)}{\left(\dfrac{\kappa}{\eta}\right)S_m^0(\chi)} = \frac{T_m(\chi,\upsilon)}{S_m^0(\chi)}\eta > \frac{T_0}{S_0} = \eta \tag{32}$$

$$T_m(\chi,\upsilon) = \max\{T^0{}_m(\chi), T^1{}_m(\chi,\upsilon)\} \tag{33}$$

Here $T_m(\chi,\upsilon)$ is maximum of two normalised tensile stresses $T^0{}_m(\chi)$ and $T^1{}_m(\chi,\upsilon)$; $S^0{}_m(\chi)$ is the maximum of the dimensionless shear stress in the middle of the beam. We introduce the breakage regime function $g(\chi,\upsilon)$, which is the ratio of the stress maxima $T_m(\chi,\upsilon)$ and $S^0{}_m(\chi)$. If a state point in plane $(\chi,\eta)$ is located above the curve $g(\chi,\upsilon)$, the breakage occurs by tensile stress; otherwise it is shear stress that causes the breakage.

To further clarify the meaning of this equality, consider a system at the point of shear failure, i.e. $S_m^0 = S_0$. If $g(\chi,\upsilon)<\eta$, then it follows that $T_m<1$. Thus, at the point of shear failure, the tensile stress does not exceed the tensile strength and tensile failure is not expected. Thus, in the situation of a gradually increasing load, shear failure will occur before tensile failure. Similarly, if $g(\chi,v)>\eta$, then $T_m>1$ (for $S_m^0=1$) and thus at the onset of shear failure, the condition for tensile failure is already satisfied and therefore the particle will first experience tensile failure.

The tensile-stress diagram shows that as $\chi$ changes from zero to $\chi_m$, different sequences of domains appear in three intervals of Poisson's ratio: $[0,v_1]$, $[v_1,v_2]$, and $[v_2,0.5]$ (Fig. 7a). A graph of the breakage function $g(\chi)$ is presented in Figs. 11a, d, and g in those three intervals, respectively. Sharp transitions in the behaviour of the function are observed when different boundaries in the tensile-stress diagram are reached. Figs. 11b, e, and h show the plots of the numerator and denominator of expression (32) of $g(\chi,v)$ for three values of Poisson's ratio $v$, taken from the three above-mentioned intervals, by red, black, and green curves, respectively. Points $X_m$ where breakage occurs, are shown in Figs. 11c, f, and i.

The graph of function $g(\chi,v)$ allows determining whether breakage occurs by tensile or shear stress. Thus, $g(\chi,v)$ is called the breakage regime function, and plane $(\chi,v)$ with different domains – the tensile-shear diagram (Figs. 11a, 11d, 11g).

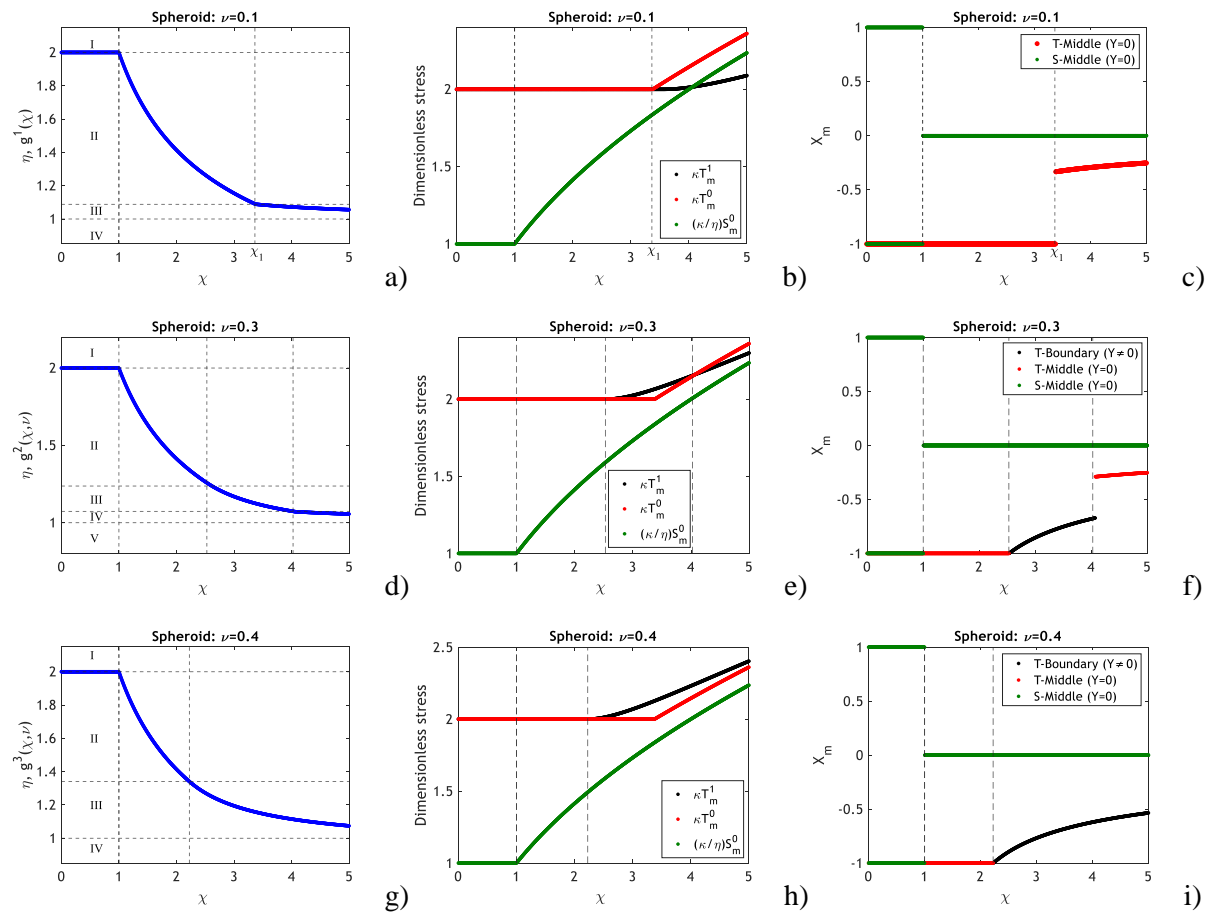

Fig. 11. Breakage regime function g ($\chi$) for different Poisson ratios $v$: a,b,c) $v<v_1$; d,e,f) $v_1<v<v_2$; g,h,i) $v_2<v<0.5$

For values of $\chi$ less than 1, the curve $g(\chi,v)$ is equal to two (Figs. 11a, d, g), and both the tensile and shear stresses are constant (Figs. 11b, e, h). Breakage occurs in the advancing point, $X_m=-1$ (Figs. 11c, f, i). As $\chi$ increases, a decrease in $g(\chi,v)$ is observed, defined by a constant tensile stress, but increasing shear stress. Shear failure for values of $\chi>1$ occurs in the particle centre, $X_m=0$. Further increases to the shape-Poisson number, $\chi$, result in an increasing tensile stress, leading to a less sharply decreasing breakage regime function. This transition occurs for only one of the two tensile stresses, while the other begins increasing at higher $\chi$. For intermediate Poisson's ratio ($v_1<v<v_2$), tensile failure occurs first on the boundary, then at higher $\chi$ it occurs in the middle of the particle (Fig. 11e). In all cases, increasing tensile stress results in tensile failure occurring at an intermediate point in the advancing half of the particle ($-1<X_m<0$, see Figs. 11e, f, i).

The value of the shape-Poisson number $\chi$ that corresponds to a given strength ratio $\eta$

$$\chi_{cr} = g^{-1}(\eta), \ g(\chi_{cr}) = \eta \tag{34}$$

is called the critical value.

For a given value of the strength ratio, $\eta$, let us discuss the determination of the failure type. In the case where the strength ratio exceeds 2 (region I in Figs. 11a, d, g), $\eta>2$, the breakage is by shear for all values of the shape-Poisson number $\chi$. Similarly, if the strength ratio is less than one, $0<\eta<1$ (region IV, Figs. 11a, g, or region V for $v_1<v<v_2$, Fig. 11d), the breakage is by tensile failure for all values of the shape-Poisson number $\chi$. For values of $\eta$ between 1 and 2 (regions II-III, Figs. 11a, d, g, and region IV for $v_1<v<v_2$, Fig. 11d), breakage will occur by tensile failure for values of $\chi$ less than the critical value, $\chi_{cr}$, and by shear failure for values larger than it.

## 4. Detrital fines detachment against electrostatic DLVO forces

Following Derjaguin and Landau 1941, Verwey and Overbeek, 1948, Israelachvili 2015, Bradford et al. 2013, this section briefly presents the fines detachment theory for detrital fines.
13, 21, 78, 79

Detrital fines have been brought to the rock by groundwater flows after being broken-off the rock surface and attached to the rock surface by electrostatic forces. Figs. 2b, 2c show the forces exerting on an attached detrital particle: attaching electrostatic $F_e$ and gravity $F_g$ forces, and detaching drag $F_d$ and lift $F_l$ forces. Electrostatic is a potential force, where the energy potential $E$ depends on the particle-surface separation distance, $h$:

$$F_e(h) = -\frac{\partial E(h)}{\partial h} \tag{35}$$

An energy profile with only a single minimum has an inflection point $h=h_m$, where the electrostatic force reaches its maximum:

$$\frac{\partial F_e(h_m)}{\partial h} = -\frac{\partial^2 E(h_m)}{\partial h^2} = 0, \ \frac{\partial^2 F_e(h_m)}{\partial h^2} = -\frac{\partial^3 E(h_m)}{\partial h^3} < 0 \tag{36}$$

Mechanical equilibrium of a particle on the rock surface in the case of favourable attachment (one primary energy minimum) is determined by the following conditions: equality of detaching and attaching torques

$$M_b = 6\pi\mu_f r_s V f_d(\alpha_s) b f_M(\alpha_s) = F_e(h) l_n, \tag{37}$$

equality of detaching and attaching force projections on horizontal

$$F_d = 6\pi\mu_f r_s V f_d(\alpha_s) = v_c F_e(h), \tag{38}$$

and equality of detaching and attaching force projections on vertical

$$F_l(U) = F_e(h) + F_g \tag{39}$$

where $l_d$ and $l_n$ are the lever arms for drag and normal forces, respectively, and $v_C$ is the Coulomb friction coefficient. Particle detachment occurs when the left hand side of (37), (38), or (39) exceeds the right hand side when $h=h_m$ (at the maximum electrostatic force). When these terms are insufficient to detach the particle, they are equilibrated by a smaller electrostatic force, and the particle sits at some distance $h<h_m$. Breach of either equilibrium conditions (37), (38), or (39) yields fines detachment by rolling, sliding, and lifting, respectively. In the case of two energy minima, there are two separating distances $h=h_m$ that correspond to both energy minima, and the equilibrium conditions (37), (39), and (39) are applicable to the particles that are located in primary and secondary energy minima. It is implicitly assumed that the particles detached by the three criteria (37), (39), and (39) continue rolling over the rock surface, sliding over the surface, and move off the surface into the liquid stream, respectively.

Under dominance of electrostatic force in Eq. (39), lifting does not occur, and the lifting criterium can be dropped.

**5. Maximum retention function as a mathematical model for fines detachment**

This section develops a novel model for fines detachment by breakage that includes a derivation of breakage flow velocity (section 5.1), model for breakage detachment and its expression as a maximum retention function MRF (section 5.2), and determination of the breakage parameters from and experimentally derived MRF (section 5.3).

*5.1. Determination of breakage velocity*

Let us determine the breakage velocity based on either formula for normalised stresses (22, 24, 29) or the *($\chi,\upsilon$)* and *($\chi,\eta$)* diagrams.

First, we compare normalised tensile stresses in the middle and on the boundary, using Eqs. (22) and (24). Since both normalised stresses are proportional to $\kappa^{-1}$, the choice of maximum normalised tensile stress is determined by the values in brackets in Eqs. (22, 24) (dimensionless stresses), so the knowledge of $\kappa^{-1}$ is not required at this stage. The choice between $T^0$ and $T^1$ can be done using the *($\chi,\upsilon$)* diagram (Fig. 7a).

Then, we compare the maximum normalised tensile stress $T_m$ with the normalised shear stress in the middle $S^0_m$, using Eqs. (22, 24) and (29). Also, both normalised stresses are proportional to $\kappa^{-1}$, so the choice is based on the comparison between the value in brackets in Eq. (22, 24) and the value in brackets in Eq. (29) times the strength ratio. As an alternative, the choice between maximum normalised tensile stress and normalised shear can be done using breakage function *$\eta = g(\chi,\upsilon)$* in *($\chi,\eta$)* plane (Fig. 11).

So, at this stage the breakage regime was determined. Consider the chosen maximum stress equation, either (22), (24), or (29). At the point of failure, the chosen normalised stress is equal to one, the value in bracket has already been calculated, allowing us to determine the strength-drag number $\kappa$. For cases where the maximum is tensile stress in the middle, tensile stress on the boundary, and shear stress in the middle, the formulae for strength-drag numbers $\kappa$ are:

$$\kappa = \begin{cases} 2, & \chi \leq \chi_1 \\ \sqrt{-\frac{\sqrt{(4\chi-1)(4\chi-9)}-4\chi+3}{8\chi}} + \sqrt{\left(\sqrt{-\frac{\sqrt{(4\chi-1)(4\chi-9)}-4\chi+3}{8\chi}}\right)^2 + \chi\left(1-\left(\sqrt{-\frac{\sqrt{(4\chi-1)(4\chi-9)}-4\chi+3}{8\chi}}\right)^2\right)^2}, & \chi > \chi_1 \end{cases}$$
(40)

$$\kappa = \begin{cases} 2, & \xi \leq 2 \\ \xi(\xi-1)^{-0.5}, & \xi > 2 \end{cases} ; \quad \xi = \chi \frac{4(1+2\upsilon)^2}{(3+2\upsilon)^2}$$
(41)

$$\kappa = \frac{T_0}{S_0} \begin{cases} 1, & \chi < 1 \\ \sqrt{\chi}, & \chi > 1 \end{cases},$$
(42)

respectively.

Substituting expression for drag from Eq. (1) into Eq. (12), we obtain

$$\kappa^{-1} = \frac{1}{2T_0}\frac{F_d}{I}r_b^2 \frac{bf_M}{r_b} = \frac{1}{T_0}\frac{2F_d}{\pi r_b^2}\frac{\alpha_s f_M}{\delta} = \frac{12}{T_0}\frac{\pi\mu_f r_s V f_d}{\pi r_b^2}\frac{\alpha_s f_M}{\delta} = \frac{12}{T_0}\frac{\pi\mu_f r_s U f_d}{\pi r_b^2}\frac{\alpha_s f_M}{\delta\phi} = \frac{12\alpha_s f_M f_d}{\delta\phi}\frac{\mu_f r_s U}{T_0 r_b^2}$$
(43)

This allows determining the breakage velocity for authigenic particles

$$U_{cr}^b = \frac{\delta\phi T_0 r_b^2}{12\alpha_s f_M f_d \mu_f r_s}\kappa^{-1}$$
(44)

where $\kappa$ is calculated by either of three formulae (40), (41), or (42).

Critical breakage velocity versus spheroidal particle radius for different aspect ratios is investigated in Fig. 12a. The smaller is the particle, the higher is the breakage velocity. So, in a test with piecewise constant increasing velocity, the largest particles are detached first, and further detachment is continuing in the order of decreasing particle size.

However, critical breakage velocity $U^b_{cr}$ is non-monotonic with respect to aspect ratio – the two smallest critical velocities are exhibited by very oblate spheroids ($\alpha_s$=0.025) and for perfect spheres ($\alpha_s$=1) . Fig. 12b shows that maxima of breakage velocity are reached for intermediate aspect ratios. The effect is attributed to non-monotonicity of the product of three $\alpha_s$ dependent functions in the denominator of the expression (44) (Fig. 12c). The overall drag is the total of surface integrals of pressure gradient and viscous shear over the particle surface. The flatter is the particle, the lower is the aspect ratio, the lower is the particle cross-section transversal to flow. This results in a lower pressure-gradient component of drag, but a higher shear viscosity component due to an increase of the contact area and the Couette flow alignment. So, the non-monotonicity of drag shown in Fig. 12d is the result of two effects of pressure gradient and shear, i.e., normal and tangential components of drag, which are competing and at odds with each other. The above also explains non-monotonicity of lift, shown in the same figure.

Fig. 12e shows that the higher is the aspect ratio, the lower are drag, moment and lift factors. Lift is significantly lower than the drag; both forces are non-monotonic $\alpha_s$-functions (Fig. 12e).

For the case of long cylinders, all the dependencies, calculated above for spheroidal particles, become monotonic for long cylinders (Fig. 13). The larger the particle the lower is the breakage velocity, i.e., the long cylinders are detached in order of decreasing of their sizes; large particles are broken first during velocity increase (Fig. 13a). The higher is the aspect ratio, the lower is the breakage velocity, i.e., it is easier to break a long thin cylinder (Fig. 13b). The aspect-ratio-dependent group that enters the expression (44) is monotonic too (Fig. 13c). The higher is the aspect ratio, the higher are the drag force (Fig. 13d) and drag factor (Fig. 13e). The lift is negligible.

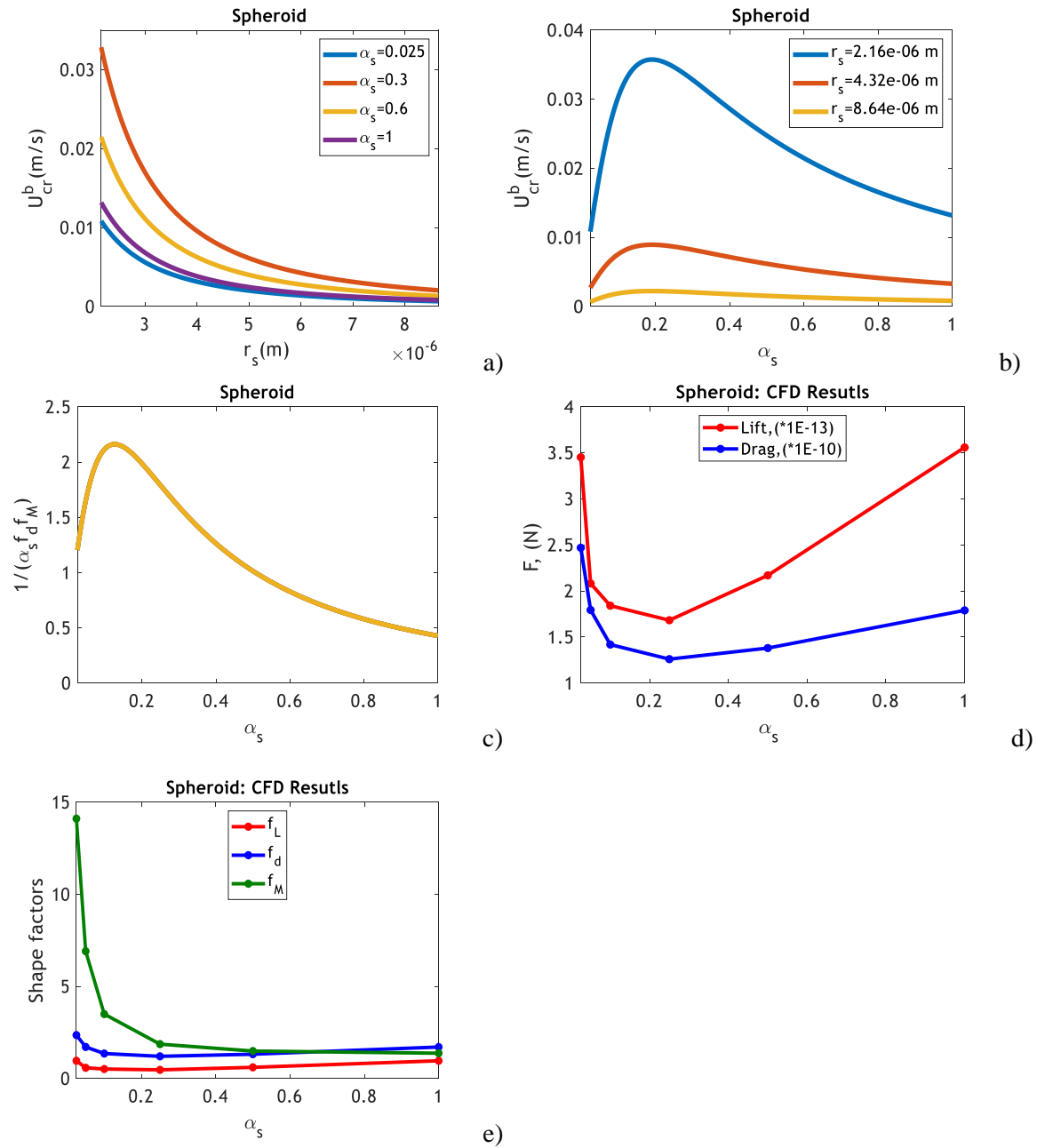

Fig. 12. Critical breakage velocity of oblate spheroids: a) Dependence of critical breakage velocity on particle radius; b) Dependence of critical breakage velocity on particle aspect ratio; c) ; d) aspect-ratio dependency for shape factor; e) aspect-ratio dependencies for drag and lift.

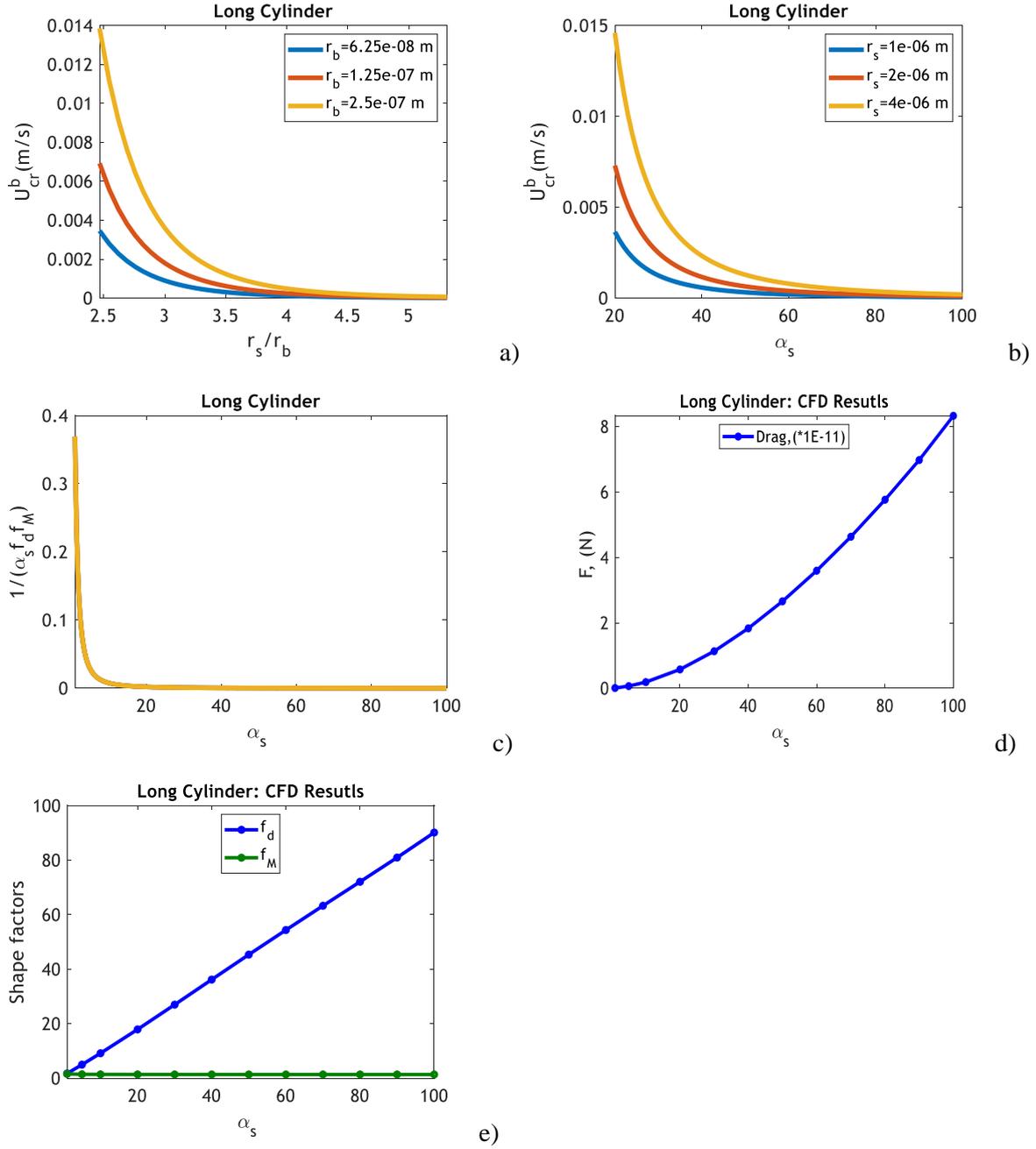

Fig. 13. Critical breakage velocity of long cylinders: a) Dependence of critical breakage velocity on particle radius; b) Dependence of critical breakage velocity on particle aspect ratio; c) c) dependency of $\alpha_s f_d f_M$ of $\alpha_s$; d) aspect-ratio dependency for shape factor; e) aspect-ratio dependencies for drag and lift.

*5.2. Formulation of fines breakage model: maximum retention function*

Breakage criterion (32, 33) yields the expression (44) of the breakage velocity $U^b_{cr}$ versus particle and rock surface properties. For uniform plane substrate and the attached particles, all the particles remain attached for velocities lower than $U^b_{cr}$, and all become detached for any velocity above $U^b_{cr}$. However, fines detachment during the velocity increase occurs gradually.

Consider a multidimensional manifold of particles of different sizes and forms situated at various sites of an asperous, micro heterogeneous rock surface. Flow velocity of creeping flow near the rock surface is probabilistically distributed over the porous space. Under the assumption of Stoke s flow in the porous space, velocity distribution over the porous space is determined by the macroscale Darcy's velocity U, i.e., local microscale speeds are proportional to $U$.[80] The mechanical equilibrium failure conditions (32, 33) indicate whether each attached to rock surface particle is broken or remains attached under a given velocity $U$. It makes attached concentration a function of velocity that is called the MRF (maximum retention function). The MRF $\sigma_{cr}(U)$ is a mathematical model for particle mobilisation by breakage. For 3D flows, the MRF is a function of the modulus of velocity:

$$\sigma_a = \sigma_{cr}^b(|U|) \tag{45}$$

Eq. (44) contains coefficients reflecting properties of authigenic particles, the particle-substrate bonds, and porous medium: tensile strength $T_0$, aspect ratio $\alpha_s$, fluid viscosity $\mu_f$, particle radius $r_s$,

Poisson ratio $\upsilon$, bond radius $r_b$, bond ratio $\delta$, and porosity $\phi$. These parameters determine the functional expression (45) for the MRF.

As it follows from the definition of the MRF as the total concentration of particles that remain attached under a given velocity $U$, the MRF monotonically decreases from zero velocity and tends to zero as velocity tends to infinity. A minimum breakage velocity, where the "first" particle is broken off, corresponds to equality of the MRF $\sigma_{cr}(U_{min})$ to the initial concentration of movable particles. Similarly, for an MRF which reaches zero at a finite velocity, this maximum breakage velocity, $U_{max}$, corresponds to the minimum flow velocity that yields failure for all bonds over the rock surface.

*5.3. Maximum retention function for simultaneous breakage and DLVO detachment*

Eqs. 1, 2, and 3 as applied with the local flow velocity around the attached particle, allow determining whether any arbitrary particle is detached for any velocity $U$, or remains attached. The maximum retention function for DLVO attraction, $\sigma^e_{cr}(U)$, is determined by the total particle concentration that remain attached by DLVO forces for a given velocity $U$.[22, 23, 81, 82] Expressing the interstitial speed $V$ in Eqs. (37, 38) via Darcy's velocity $U=V\phi$, we obtain two expressions for critical velocity for particle detachment against electrostatic DLVO forces for the conditions of rolling and sliding:

$$U_{cr}^e = \frac{F_e \phi l_n}{6\pi \mu_f r_s f_d(\alpha_s) b f_M(\alpha_s)} \tag{46}$$

$$U_{cr}^e = \frac{v_C F_e(h)\phi}{6\pi \mu_f r_s f_d(\alpha_s)} \tag{47}$$

respectively. Here $U_{cr}^e$ is the Darcy' velocity that detaches the particles under the conditions given in right side of Eqs. (46) and (47).

Critical detachment velocity against DLVO forces is determined by the minimum critical velocity from those given by two criteria (46) and (47). Two failure criteria – by rolling and by Coulomb's friction, given by Eqs. (46) and (47), respectively, are mathematically equivalent, provided that

$$v_C = \frac{l_n}{b f_M(\alpha_s)} \tag{48}$$

Eqs. (37, 38) contain a set of parameters, which are properties of either particles or surface: particle radius $r_s$, its aspect ratio $\alpha_s$, zeta potentials of particles $\psi_{01}$ and rock $\psi_{02}$, brine salinity, pH, lever arm ratio $l$, temperature $T$, tensile strength $T_0$, and beam radius $r_b$. Continuity of the MRF $\sigma^e_{cr}(U)$ – gradual fines detachment with continuous flow velocity increase – is determined by the probabilistic distributions of those parameters.[83]

The assumption of independent detachment of particles against DLVO forces and by breakage determines the overall MRF as a sum of the two individual MRFs:

$$\sigma_{cr}(U) = \sigma^e_{cr}(U) + \sigma^b_{cr}(U) \tag{49}$$

Fig. 14 shows that the MRF is monotonically decreasing – the higher is the velocity, the lower is the attached concentration. A velocity increase from $U_n$ to $U_{n+1}$ yields detachment of particles with retained concentration $\Delta\sigma_n$:

$$\Delta\sigma_n = \sigma_{cr}(U_n) - \sigma_{cr}(U_{n+1}), \tag{50}$$

which are transformed into suspension concentration

$$c = \Delta\sigma / \phi \tag{51}$$

Fig. 14 presents the form of MRF for fines detachment against electrostatic force, by breakage, and the total MRF. Lower dashed curves represent the MRF for detrital particles, the combined dashed-solid curves correspond to the breakage MRF, and the solid curves show the total MRF.

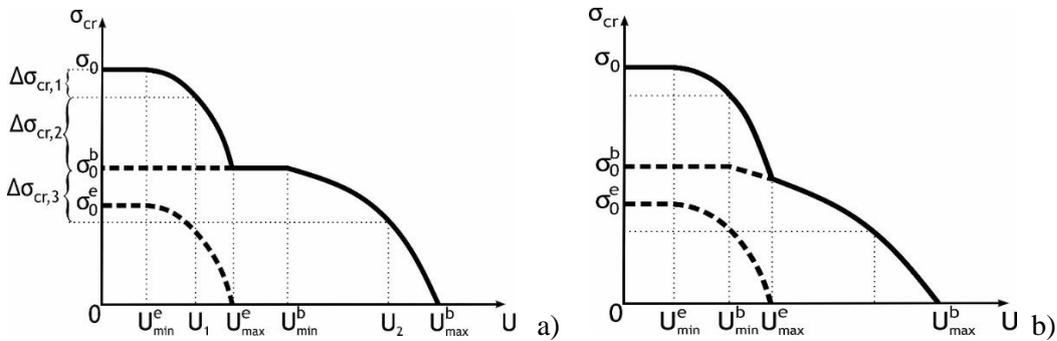

Figure 14: Total MRF for detachment of detrital and authigenic fines: a) in the case of $U^e_{max} < U^b_{min}$, MRF has plateau; b) MRF is monotonically decreasing function of velocity in the case where $U^e_{max} > U^b_{min}$

Consider piecewise-constant injection rate increase; each constant-rate stage is maintained until stabilisation. The case of Fig. 14a corresponds to weak electrostatic attraction (using low-salinity, high-pH or high-temperature water) and high particle-rock-bond strength (highly consolidated sandstones). The detrital particles start detaching at velocity $U^e_{min}$ and continue mobilisation until velocity $U^e_{max}$, where all movable detrital fines are detached. At a velocity that is higher than the minimum plateau velocity, the maximum DLVO force cannot secure any detrital particle on the rock surface, i.e., all detrital particles detach. In this case, where $U^e_{max} < U^b_{min}$, no further particle detachment occurs until the velocity reaches $U^b_{min}$, where the first authigenic particle experiences failure. All movable authigenic particles are detached once the velocity reaches $U^b{}_{max}$. In this case, the total MRF has a plateau, where all detrital particles are detached under velocities that are lower than the minimum plateau velocity, and authigenic fines remain attached until a velocity that is higher than the maximum plateau velocity. No detachment occurs at plateau velocities. For high electrostatic attraction and low strength, the velocity intervals for detachment by both causes overlap (Fig. 14b).

Existence of two plateaus in MRF indicates three detachment mechanisms. For example, if the DLVO energy profile has two minima in the case of unfavourable fines attachment and the particle-rock bond strength is high, the MRF can have two plateaus.

Finally, there appears the following *algorithm of fines detachment modelling*: (i) identifying the appropriate domain in plane $(\chi,\upsilon)$ – the tensile-stress diagram – and determining which tensile stress maximum on the beam base – along the symmetry axis or the boundary of the beam base – is higher; (ii) determining whether the failure occurs by tensile or shear stress from the inequality for the breakage regime function $g(\chi,\upsilon)>\eta$ in the shear-tensile diagram; (iii) calculation of breakage velocity from the drag-strength number $\kappa$; (iv) determining the probabilistic distribution of breakage velocity; (v) calculation of the MRF for breakage; (vi) adding the MRFs for breakage and detachment against electrostatic attraction.

MRF (49) in set of points $U_1<U_2<...<U_n$ is determined from a laboratory coreflood with piecewise-constant injection rate increase from breakthrough particle concentrations.

## 6. Laboratory study of fines detachment by breakage by coreflooding

*6.1. Laboratory study*

The laboratory study conducted as part of this study comprised of core drying, core saturation by water with 0.6 M of NaCl under vacuum, and injection of this water into the core. A Castlegate sandstone core with permeability 917 mD and porosity 0.24, length 5.1 cm and pore volume 13.7 cu cm was used in the test. Seven flow rates were applied in order of increasing rate. Fig. 15 presents the schematic of laboratory set-up. The essential parts of the setup are coreholder with the core placed in a Viton sleeve, four differential pressure transmitters, HPLC pump, effluent fraction collector, and particle counter. The overburden pressure was created by a manual pressure generator. A data acquisition module along with signal convertor provided results which were visualised on a PC screen in real time. Breakthrough concentration (Fig.16a) and size distributions of produced particles along with pressure drop across the core (Fig. 16b) have been measured during each constant-rate injection step.

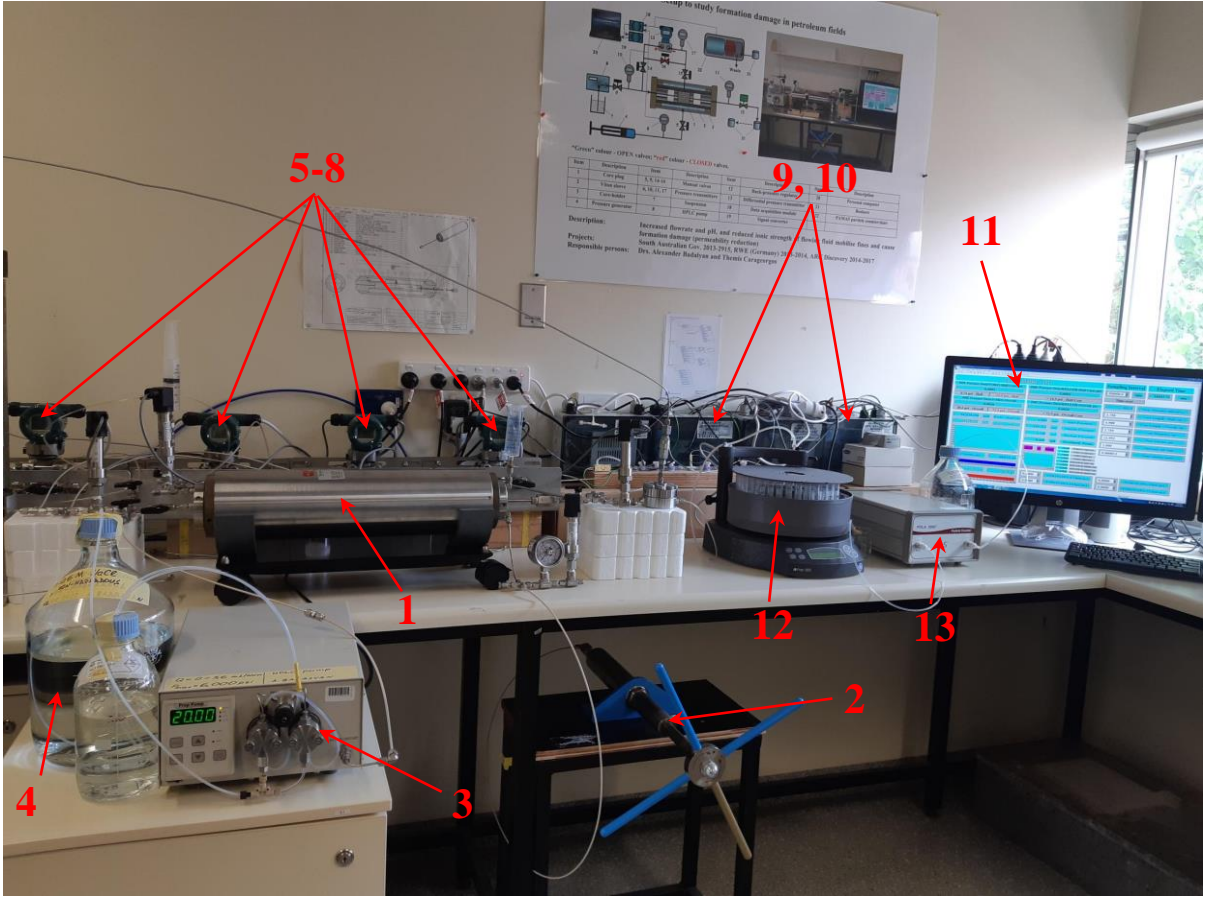

Figure 15. Laboratory set-up for particle detachment due to DLVO and breakage: 1 - core holder, 2 - manual pressure generator, 3 - high-performance liquid chromatography (HPLC) pump, 4 - brine solution, 5-8 - differential pressure transmitters, 9 – data acquisition module, 10 – signal converter, 11 - personal computer, 12 - fraction collector, 13 - particle counter.

Fig. 16a shows that at low rate, the breakthrough concentration declines exponentially, which is typical for deep bed filtration of low-concentration colloids with constant filtration coefficient. [20, 84] At high rates, this behaviour is combined with a sharp and fast concentration decrease down to low values, indicating that particle capture intensity is decreasing, until the point where all capture vacancies are filled and no further capture occurs. Here the capture rate is described by the Langmuir filtration function. [85, 86]

Relative change of average core permeability is captured by impedance $J$

$$J(t) = \frac{\Delta p(t)}{\Delta p(0)} \frac{U(0)}{U(t)} \qquad (52)$$

where $\Delta p$ is the pressure drop across the core.

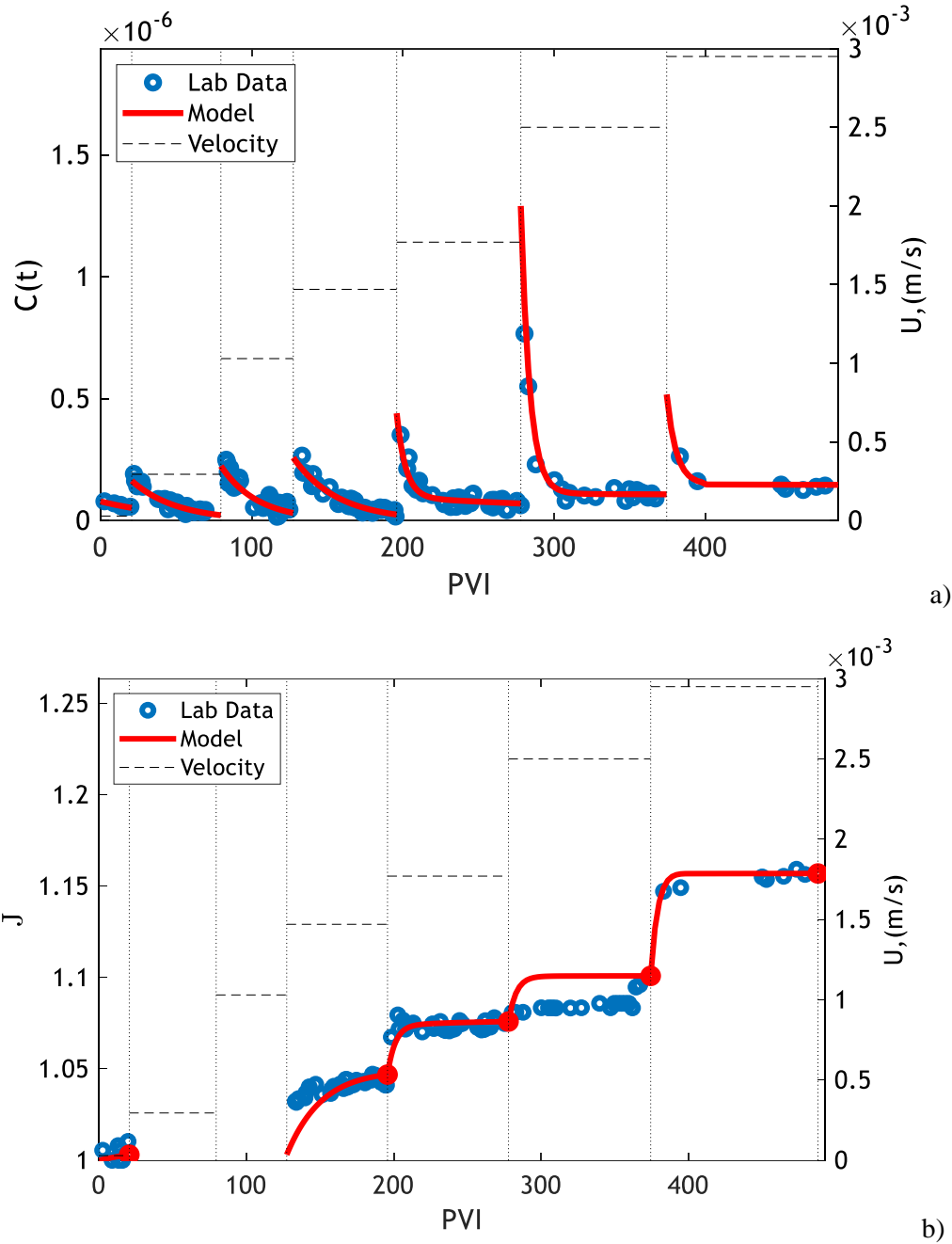

Fig. 16. Breakthrough concentrations (left axes) during the test with seven rates (right axis)

For large rates, impedance first increases sharply and then switches to slow growth (Fig. 16b). This transition occurs at approximately the same time as the suspended concentration switches from a sharp decrease to a slow decrease (Fig. 16a). Colloidal flow with a Langmuir filtration coefficient is attributed to asperous authigenic fines with rough surfaces after breakage, which are strained in the large pore throats with fast filling of all of them, while the capture of smoother detrital particles is less intensive and continues significantly longer.

The size distributions of produced particles at high rates have a clear bimodal structure, which also supports the two-population hypothesis.

The next section supports the two-population hypothesis by successful matching of experimental data with a two-population colloidal transport model.

*6.2. Mathematical model for transport of detached colloids*

To match the breakthrough and impedance histories, we have developed a two-population model of deep bed filtration, given by Eqs. (B1-B4) and described in Appendix B. The system contains mass conservations for both populations, Eqs. (B1) for $k=1,2$, where the delay of each particle population's velocity comparing with the carrier water velocity is expressed by drift-delay factors $\alpha_k$. Particle capture rates for each particle population are proportional to the respective particle fluxes $c_k\alpha_k U$; the proportionality coefficients are $\lambda_1$ and $\lambda_2$. The filtration coefficients $\lambda_1$ and $\lambda_2$ are equal to probabilities of the particle capture by the rock per unit length of the particle trajectory in the porous space. The first population filtrates with constant filtration coefficient $\lambda_1$, and the filtration coefficient for the second population $\lambda_2(\sigma_2)$ has Langmuir form given by Eq. (B4). The interaction of two population fluxes is expressed by the joint contribution of the retained particles of both types to the overall rock hydraulic resistance. Dependence of permeability on both retained concentrations in Eq. (B3) is obtained by first order Taylor's expansion of the inverse to permeability as a function of the retained concentrations of two populations. Here $k_0$ is the initial (undamaged) permeability, and formation damage coefficients $\beta_1$ and $\beta_2$ quantify the extent of permeability decline caused by each of the particle populations.

The assumption of the independence of the model parameters of fluid pressure separates four equations (B1, B2) for $k=1,2$ from Eq. (B3). Dependence of the individual model parameters on their particular retention concentrations separates systems (B1, B2) for $k=1,2$ from each other. Initial conditions (B5) correspond to electrostatic and breakage detachment with instant particle mobilisation: the detached concentration $\Delta\sigma_{cr}$ due to change of flow rate from $U_{n-1}$ to $U_n$, determined by MRF is instantly translated into suspension concentration $\Delta\sigma_{cr}/\phi$.

Initial-boundary problem (B5, B6) for 1D flow system (B1, B2) allows for an exact solution.[86-88] Suspended and retained concentrations of both populations are expressed by explicit formulae (B7). This allows for the derivation of an explicit formula for the pressure gradient. Yet, the pressure drop across the core and impedance are calculated by numerical integration.

*6.3. Results*

Fig.16 shows the experimental data by blue dots and the matched modelling data by red curves. High match for breakthrough concentrations is supported by the coefficient of determination $R^2=0.96$. Breakthrough concentrations highly exceed the accuracy of the particle counter used in the set-up (Fig. 15). The accuracy of pressure transducers exceeds the measured pressure drops only by a factor of 2-3, so the formation damage coefficients have been tuned by the final impedance values alone, which are shown by red dots in Fig. 16b. However, the coefficient of determination for the overall impedance is also high $R^2=0.94$.

The parameters obtained from tuning − filtration coefficients $\lambda_1=98$ 1/m and $\lambda_2=87$ 1/m, formation damage coefficients $\beta_1=5300$ and $\beta_2=7300$, and delay-drift factors $\alpha_1=5\times10^{-3}$ and $\alpha_2=23\times10^{-3}$ − belong to commonly reported intervals.[13, 20, 30, 84] The tensile strength for kaolinite is determined from Eq. (44), which presents an unbiased breakage velocity estimate, and is equal to $T_0=0.15$ MPa. Although this value is lower than typical tensile strengths for minerals, significantly smaller values for coals and clays have been reported.[31, 32, 48, 89, 90] The detached concentration $\Delta\sigma_{cr}(U)$ versus velocity increases at small rates from zero and declines at high velocities up to zero, which complies with the typical form of the maximum retention function curve.[22, 23, 81, 91] High match shown in Fig. 16 validates the two-population model for deep bed filtration of broken-off and DLVO-detached fines.

Close match by a single-population model has been achieved for lab data on coal flooding by Guo 2016 and Huang et al. 2017, 2021, and on sandstone floods by Ochi and Vernoux 1998, and Torkzaban et al. 2015.[30, 73, 92-94] Here the coefficient of determination $R^2$ varies in the interval [0.88,0.94], exhibiting a close match. The tuned parameters have the same order of magnitude as those obtained in our test.

## 7. Feasibility of fines breakage during injection and production in natural reservoirs

A fines migration test is a routine laboratory experiment to prevent formation damage due to fines migration. A core is submitted to flow with piecewise-constant increasing velocity. Minimum fines migration velocity is determined by appearance of the first fine particles in the effluent. This determines maximum well rate under which fines migration does not occur. Here we predict the minimum velocity for fines migration using the stress diagram technique and conclude about the viability of fines detachment by breakage during water, oil, and gas injection and production. The results below also illustrate using the sequential techniques for tensile-stress diagram, shear-stress diagram, and equation (44) for the breakage velocity.

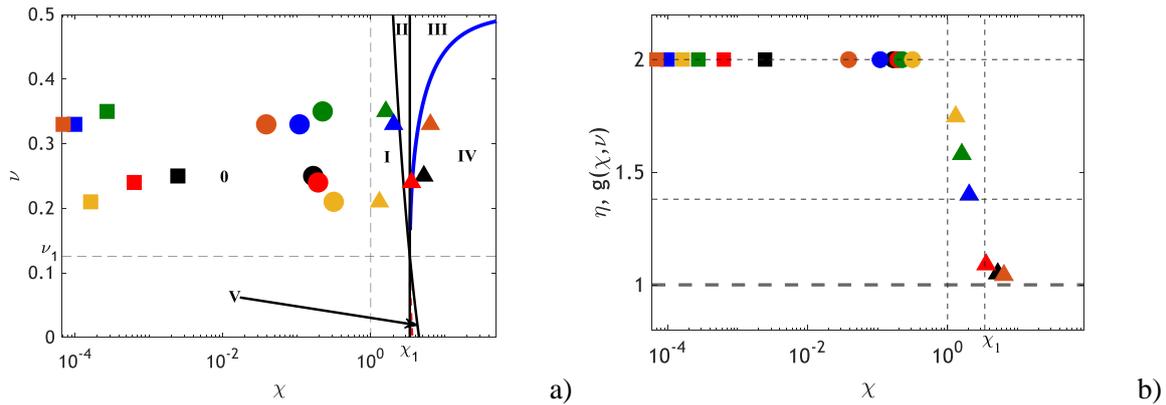

Fig. 17. Determining breakage regime for 6 examples of injection and injection in natural reservoirs: a) tensile-stress diagram, b) shear-tensile diagram

The data for the cases of production of heavy oil, polymer injection, dewatering of coal seams, $CO_2$ injection, well fracturing by water, and well fracturing by highly-viscous fluid, which are marked in Fig. 17 by black, red, green, yellow, blue, and orange, respectively, are taken from the corresponding papers by Ado 2021, Gao 2011, Shi et al. 2008, Spivak et al. 1989, and Prasetio et al. 2021.[95-99] Circular, triangular, and square points in Fig. 17 correspond to spheroidal, flat and long cylinder particles. Volume-equivalent particle size $r_s=2\times10^{-6}$ m and wellbore radius $r_w=0.1$ m are assumed for all cases. Poisson's ratios $\upsilon$ varies in the interval 0.21 to 0.35. Fluid viscosities were taken as high as 1000, 300, and 40 cp for fracturing fluid, heavy oil, and polymer solution and as low as 0.02 cp for $CO_2$. For the composite particle-rock bond, a low value of tensile strength $T_0=0.2$ MPa is assumed for all cases; yet several works report lower $T_0$ values.[31, 32, 45, 89] Other data are presented in the Table 1. The values of shape-Poisson numbers $\chi$ have been calculated by Eq. (12), which allows placing the corresponding state points into the tensile-stress diagram ($\chi,\upsilon$) in Fig. 17a. Values of $\chi$ and $\upsilon$ allow calculating breakage function $\eta=g(\chi, \upsilon)$ for the state points and determine their position in the tensile-shear diagram ($\chi,\eta$) (Fig. 17b). Here the state points are located on breakage regime curves $\eta=g(\chi,\upsilon)$ that correspond to different Poisson's ratios. Breakage function $g(\chi, \upsilon)=2$ for spheroidal and long cylindrical cases. For all cases, like in papers by Gao et al. 2014, Guo et al. 2018, and Horabik and Jozefaciuk 2021, the strength ratio $\eta$ is assumed to be equal to one.[100-102] As discussed earlier, for values of $\eta$ lower than or equal to 1, breakage occurs by tensile failure (Fig. 11).

Table 1: Calculation of minimum fines breakage velocity of the fluid at the well / fracture wall

|  | $\alpha_s$ | $\delta$ | $\upsilon$ | $\chi$ | $\mu$, (Pa.s) | $U_{cr}^b$, (m/s) |
|---|---|---|---|---|---|---|
| Heavy oil | 1 | 0.8 | 0.25 | 0.17 | 0.30 | $4.14 \times 10^{-4}$ |
| Polymer | 0.8 | 0.7 | 0.24 | 0.19 | 0.04 | $2.89 \times 10^{-3}$ |
| CBM | 0.6 | 0.6 | 0.35 | 0.22 | $1 \times 10^{-3}$ | 0.17 |
| $CO_2$ | 0.4 | 0.5 | 0.21 | 0.32 | $2.28 \times 10^{-5}$ | 3.02 |
| HF-W | 0.2 | 0.4 | 0.33 | 0.11 | $1 \times 10^{-3}$ | $7.74 \times 10^{-2}$ |
| HF | 0.1 | 0.3 | 0.33 | 0.04 | 1 | $1.64 \times 10^{-5}$ |

Fig. 17a shows that for spheroidal and long cylindrical particles, all 12 cases belong to domain 0, where both tensile stresses $\kappa T^0_m$ and $\kappa T^1_m$ are equal to two, while for short cylindrical particles some points belong to domain I, III, IV. For both cases of maximum tensile stress in the middle and the boundary of the beam base, Eqs. (40) and (41) show that the strength-drag numbers are equal, i.e. $\kappa=2$. It allows calculating minimum breakage velocities using Eq. (44). Points in Fig. 17b correspond to different Poisson's ratios and, therefore, are located on different breakage regime curves, $\eta=g(\chi,\upsilon)$. However, the overall curve has the same general tendency as those presented in Fig. 11.

The breakage velocities are presented in the eighth column of Table 1. The breakage velocities for heavy oil production, hydraulic fracturing by water, and dewatering of coal bed seam are untypically high, i.e., fines breakage is unlikely. The calculated breakage velocity for polymer solution is feasible during injection in a highly permeable reservoir. The calculated breakage velocity for hydraulic fracturing by a highly viscous fluid is typical. The $CO_2$ breakage injection velocity is too high for field cases, but it can decrease down to any arbitrarily low value during mineral dissolution of particle-rock bond in carbonic acid. The same corresponds to waterflooding of carbonate reservoirs during rock dissolution in water. The above demonstrates likelihood of particle breakage near to injection and production wells.

## 8. Conclusions

Integration of Timoshenko's beam theory with breakage criteria for particle-substrate bonds and CFD flow modelling allows concluding the following.

Breakage conditions where either of the tensile or shear stresses reaches the strength value, is determined by three dimensionless groups: the strength-drag number $\kappa$, the aspect-Poisson number $\chi$, strength ratio $\eta$, as well as the Poisson's ratio $\upsilon$.

Stress maxima are reached at the base of the particle, either at the central axis, $Y=0$ or at the base boundary, $X^2+Y^2=1$.

Stress maxima along the axis $Y=0$ are determined by the aspect-Poisson number $\chi$ alone, while the maxima at the beam boundary are determined by both the aspect-Poisson number $\chi$ and Poisson's ratio υ. Equality of tensile stress maxima at the beam base axes and boundary separates the plane (χ,υ), which is called the tensile stress diagram, into 5 domains, where one tensile maximum exceeds the other.

Shear stress maximum at the central axis of the particle base is always higher than that at the boundary.

The breakage regime – by either tensile or shear stress – is determined by breakage function *g(χ,υ)*, which is the ratio of stress maximum of two tensile maxima and maximum shear. The breakage regime depends on the shape-Poisson number, Poisson's ratio, and the ratio η between tensile and shear strengths. The breakage occurs by tensile failure if *g(χ,υ)>η,* i.e., where the point *(χ,η)* is located below the curve *g(χ,υ)* in the shear-tensile diagram *(χ,η)*. If the strength ratio exceeds 2, particles are detached by shear stress for all values of χ and υ. For strength ratios below two and above one, the particles are detached by shear stress for shape-Poisson numbers χ such that *η* exceeds g(χ,υ); for lower aspect-Poisson ratios, particles are detached by tensile stress. For strength ratios below one, the particles are detached by tensile stress for all Poisson's ratios.

The definition of the breakage regime – by either tensile or shear stress, in the base middle or at the boundary - is independent of flow velocity. For an identified breakage regime, the breakage flow velocity is determined by the strength-drag number *κ(χ,υ)* alone. For a given particle shape, the critical breakage velocity is proportional to the strength and particle size and is inversely proportional to viscosity. These conclusions are the consequence of the assumptions of a Newtonian fluid, elastic beam deformation, and the strength failure criteria.

During bond breakage under increasing velocity, the particles of all shapes detach in order of decreasing of their radii, i.e., the large particles break first. For particles of the same volume, the breakage velocity versus aspect ratio is non-monotonic for spherical particles – very flat or almost spherical particles are detached at low flow velocities while the highest flux detaches particles with intermediate aspect ratio. However, for long cylinders, the product of drag and moment factors by the aspect ratio versus aspect ratio is a monotonically decreasing function, and the lower the aspect ratio the higher the breakage velocity.

A mathematical model for colloidal detachment by breakage is a maximum retention function (MRF), derived from the formula for breakage flow velocity.

With simultaneous detachment of authigenic and detrital fines, the MRF is the total of those obtained from mechanical equilibrium conditions for detachment against electrostatic attraction, and by breakage. For weak DLVO particle-rock attraction and high bond strength, the total MRF has a velocity plateau, where all detrital particles are already detached and authigenic particle breakage hasn't started yet. For high electrostatic attraction and low bond strength, the velocity intervals for detachment by both causes overlap, and the plateau disappears.

The breakage MRF allows closing the governing equations for migration of authigenic clays with bond failure. The total MRF for authigenic and detrital fines can be determined from breakthrough particle concentration during a coreflood with piecewise constant and increasing velocity. The determination of the MRF is based on the analytical model for fines mobilisation, migration, and straining (size exclusion). Matching the breakthrough curve allows determining the MRF along with the filtration coefficient for straining and drift-delay factor.

For an MRF with a plateau, the initial percentage of authigenic and detrital fines is calculated directly from the plateau height. For MRF without plateau, the percentage of authigenic and detrital fines is calculated by tuning the MRF coefficients.

The model for fines detachment by breakage is validated by the coreflood with 7 constant-rate injections by the two-population feature of produced particle concentrations. High match of the breakthrough concentrations and pressure drop using the two-population model, tuned coefficients within commonly reported intervals, as well as bimodel size distributions of produced fines all support the validity of the proposed formulation.

Calculations of breakage velocity shows that breakage of authigenic particles can occur during $CO_2$ injection for storage, polymer injection into oilfields, leak-off of highly viscous fracturing fluids during well fracturing, and waterflooding in carbonate oilfields. Fines detachment by breakage during heavy oil production, dewatering of coal seams, and well fracturing by water is unlikely.

**Appendix A. Expressions for stresses in 3D elastic beam model**

Under the model assumptions formulated in section 2.1, Timoshenko's 3D solution for elastic deformation of a cylindrical beam (Fig. 4) shows that the normal stress $\sigma_z$ reaches a maximum at the beam bottom, $z=0$, and the two shear stresses $\tau_{xz}$ and $\tau_{yz}$ are independent of $z$. The normal and shear stresses at the beam base with applied external load $F_d$ are: [33]

$$\sigma_z = \frac{F_d b f_M x}{I}, \quad \sigma_x = \sigma_y = \tau_{xy} = 0 \tag{A1}$$

$$\tau_{xz} = \frac{(3+2\upsilon)}{8(1+\upsilon)} \frac{F_d}{I} \left( r_b^2 - x^2 - \frac{(1-2\upsilon)}{(3+2\upsilon)} y^2 \right) \tag{A2}$$

$$\tau_{yz} = -\frac{(1+2\upsilon)}{4(1+\upsilon)} \frac{F_d xy}{I} \tag{A3}$$

where, $\sigma_z$ is the normal bending stress at contact area, $\tau_{xz}$ is the shear stress acting on the $z$ plane and towards the $x$ direction, $\tau_{yz}$ is the shear stress acting on the $z$ plane and towards the $y$ direction, $\upsilon$ the is Poisson's ratio, and $I$ is the moment of inertia, which for circular cross-section is equal to $\pi r_b^4/4$.

As expressed by Eq. A1, the normal bending stress expands the matter at $x<0$ reaching a minimum in the advance point $x=-r_b$, $y=0$ (Fig. 4a) and contracts at $x>0$ reaching a maximum at the receding point $x=r_b$, $y=0$. Fig. 4b illustrates Eq. (A2) and corresponds to shear stress that opposes the external load $F_d$ and is equal zero only in advance and receded points. Eq. (A3) is illustrated by Fig. 4c, showing the transversal shear stress.

The stress tensor, as per solution (A1-A3) is:

$$\begin{bmatrix} 0 & 0 & \tau_{xz} \\ 0 & 0 & \tau_{yz} \\ \tau_{xz} & \tau_{yz} & \sigma_z \end{bmatrix} \tag{A4}$$

The principal stresses are eigen values of the stress tensor (A4):

$$\sigma_1 = \frac{\sigma_z + \sqrt{\sigma_z^2 + 4(\tau_{xz}^2 + \tau_{yz}^2)}}{2}, \quad \sigma_2 = 0, \quad \sigma_3 = \frac{\sigma_z - \sqrt{\sigma_z^2 + 4(\tau_{xz}^2 + \tau_{yz}^2)}}{2} \tag{A5}$$

where, $\sigma_1$, $\sigma_2$, $\sigma_3$ are the principal stresses in decreasing order of magnitude, and

$$\sigma_1 > \sigma_2 = 0 > \sigma_3. \tag{A6}$$

Maximum tensile and shear stresses in Eqs. (8, 9) correspond to points $(\sigma_3, 0)$ and $(\sigma_1 + \sigma_3)/2$, $(\sigma_1 - \sigma_3)/2$ in plane $(\sigma, \tau)$. Consequently, the maximum tensile and shear stresses are

$$\max \sigma = \sigma_3, \quad \max \tau = \frac{\sigma_1 - \sigma_3}{2}, \tag{A7}$$

respectively.

## Appendix B. Two-population colloidal-suspension transport in porous media

We discuss deep bed filtration of two particle populations for detrital and authigenic fines. Mass balance and capture rate equations and Darcy's law for both populations are:[12,13,103]

$$\frac{\partial}{\partial t}(\phi c_k + \sigma_k) + \frac{\partial}{\partial x}(c_k \alpha_k U_n) = 0 \tag{B1}$$

$$\frac{\partial \sigma_k}{\partial t} = \lambda_k c_k \alpha_k U_n, \quad k = 1, 2, \quad n = 0, 1, 2... \tag{B2}$$

$$U_n = -\frac{k}{1 + \beta_1 \sigma_1 + \beta_2 \sigma_2} \frac{\partial p}{\partial x} \tag{B3}$$

where $\phi$ is the porosity, $c_k$ and $\sigma_k$ are the suspended and retained concentrations for both populations, k=1,2, $\lambda_k$ are the filtration coefficients, $\alpha_k$ are the drift-delay factors, $U_n$ is the flow velocity, $k_0$ is the initial permeability, $p$ is the pressure, and $\beta_k$ are the formation damage coefficients. Index $k$ corresponds to the two populations; index $n$ is attributed to the injection velocity at the $n$-th step of the test. The filtration coefficient for the first population is constant. The filtration coefficient for the second population is a blocking (Langmuir) function of retained second-population concentration[103]

$$\lambda_1 = const$$

$$\lambda_2(\sigma_2) = \begin{cases} \lambda_{20}\left(1 - \frac{\sigma_2}{\sigma_{20}}\right), & \sigma_2 < \sigma_{20} \\ 0, & \sigma_2 > \sigma_{20} \end{cases} \tag{B4}$$

Initial retained concentrations for both populations are equal to the concentrations of mobilised fines after a velocity increase

$$t = 0: c_{k0} = \frac{\sigma_{cr}^k(U_n) - \sigma_{cr}^k(U_{n-1})}{\phi} = \frac{\Delta \sigma_{cr}^k(U)}{\phi}, \quad k = 1, 2 \tag{B5}$$

Inlet boundary condition corresponds to the injection of particle-free water

$$x = 0 : c_k = 0, \quad k = 1, 2 \tag{B6}$$

Breakthrough concentration is a total of those for both populations. The exact solution is[88,89]

$$c(x,t) = c_1(x,t) + c_2(x,t)$$

$$c_1 = \begin{cases} \dfrac{\Delta \sigma_1(U)}{\phi} \exp\left(-\dfrac{\lambda_1 \alpha_1 U}{\phi} t\right), & x > \dfrac{\alpha_1 U}{\phi} t \\ 0, & x < \dfrac{\alpha_1 U}{\phi} t \end{cases}$$

$$c_2 = \begin{cases} 0, & x > \dfrac{\alpha_2 U}{\phi} t \\ c_2 = \dfrac{\sigma_m \left(\dfrac{\Delta \sigma_2}{\sigma_m} - 1\right)}{\phi \left\{ 1 - \dfrac{\sigma_m}{\Delta \sigma_2(U)} \exp\left[\dfrac{\lambda_2 \alpha_2 U}{\phi} t \left(1 - \dfrac{\Delta \sigma_2}{\sigma_m}\right)\right] \right\}}, & x < \dfrac{\alpha_2 U}{\phi} t \end{cases}$$

(B7)

## Acknowledgements

The authors are deeply grateful for fruitful discussions and support to A/Profs Giang D. Nguyen and Abbas Zeinijahromi, Drs Noune Melkoumian, Zhao Feng Tian, Themis Carageorgos and Heng Zheng Ting (University of Adelaide), Prof Leslie Banks-Sills (Tel-Aviv University) and A/Prof Hamid Roshan (University of New South Wales).

99. Prasetio MH, Anggraini H, Tjahjono H, Pramadana AB, Akbari A, Madyanova M, et al., editors. Success Story of Optimizing Hydraulic Fracturing Design at Alpha Low-Permeability Reservoir. SPE/IATMI Asia Pacific Oil & Gas Conference and Exhibition; 2021: OnePetro.
100. Gao F, Stead D, Kang H. Numerical investigation of the scale effect and anisotropy in the strength and deformability of coal. International Journal of Coal Geology. 2014;136:25-37.
101. Guo W-Y, Tan Y-L, Yu F-H, Zhao T-B, Hu S-C, Huang D-M, et al. Mechanical behavior of rock-coal-rock specimens with different coal thicknesses. Geomechanics and Engineering. 2018;15(4):1017-27.
102. Horabik J, Jozefaciuk G. Structure and strength of kaolinite–soil silt aggregates: Measurements and modeling. Geoderma. 2021;382:114687.
103. Bedrikovetsky P. Mathematical theory of oil and gas recovery. Springer Science & Business Media; 2013.
38